\documentclass[10pt,sigplan,letterpaper,twocolumn]{article}
\usepackage[10pt,nocopyright]{sigmin}

\usepackage{listings}
\usepackage{color}
\usepackage[hyphens]{url}                                         %%
\usepackage[breaklinks,colorlinks]{hyperref}                      %%
\hypersetup{citecolor=blue,linkcolor=blue}                        %%
\usepackage{amsmath,amsopn,amssymb,amsthm}                        %%
\usepackage{thmtools}
\usepackage{thmbox}
\usepackage[toc,page]{appendix}
\usepackage{subcaption}                                           %%
\usepackage{endnotes,microtype,xspace,fancyvrb,multirow} %%
\usepackage{epigraph} 

\usepackage{booktabs}      
\usepackage{tabularx}
\usepackage{array,underscore,relsize}                             %%
\usepackage[T1]{fontenc}                                          %%
\usepackage{times}                                                %%
\usepackage{fancyhdr,lastpage}                                    %%
\usepackage{enumitem}                                             %%
\usepackage[labelfont=bf,font=normal,skip=1pt]{caption}           %%
\usepackage[export]{adjustbox}                                    %%
\usepackage{breakurl}                                             %%
\usepackage{pifont}                                               %%
\usepackage{tablefootnote}                                        %%
\usepackage{bbding}                                        %%
\usepackage[table]{xcolor} % Allows for coloring of cells
\usepackage[linesnumbered,vlined,boxed,commentsnumbered]{algorithm2e}
\DontPrintSemicolon
\newcommand{\ra}[1]{\renewcommand{\arraystretch}{#1}}
\usepackage{amssymb}
\usepackage[normalem]{ulem}
\usepackage[hang,flushmargin]{footmisc}
\usepackage{textcomp}
\usepackage{graphicx}
\usepackage{authblk}

\usepackage[compact]{titlesec}
\titleformat*{\section}{\large\bfseries}
\titleformat*{\subsection}{\normalsize\bfseries}
\titleformat*{\subsubsection}{\normalsize\bfseries}
\titlespacing{\section}{0pt}{3ex}{1ex}
\titlespacing{\subsection}{0pt}{2ex}{1ex}
\hypersetup{
  colorlinks,
  linkcolor={red!50!black},
  citecolor={blue!50!black},
  urlcolor={blue!80!black}
}

\newcommand{\fig}[1]{Figure{~\ref{#1}}}
\newcommand{\eqa}[1]{Equation{~\ref{#1}}}

\newcommand{\cmark}{\ding{51}}%
\newcommand{\one}{\texttt{\uppercase\expandafter{\romannumeral1}}}
\newcommand{\two}{\texttt{\uppercase\expandafter{\romannumeral2}}}

\newcommand{\red}[1]{\textcolor{red}{#1}}

\newcommand{\stitle}[1]{\vspace{1.1ex}\noindent{\bf #1}}
\newcommand{\etitle}[1]{\vspace{0.8ex}\noindent{\em\underline{#1}}}

\newcommand{\highlight}[1]{\red{\textbf{#1}}}

\usepackage{tikz}
\usepackage{colortbl}
\usetikzlibrary{tikzmark,calc}

\usepackage{balance}
\newcolumntype{P}[1]{>{\centering\arraybackslash}p{#1}}

\definecolor{appcolor}{RGB}{191,255,255}
\newcommand\PartiallyFilledCellWidth[4]{
\tikz[overlay,remember picture] 
    \fill[#1] ( $ (pic cs:#2) + (-0.1pt,-0.1pt) $ ) rectangle ++(#4*#3cm - 0.8pt,1.8ex);
}

\begin{document}

%\title{\Large \bf Characterizing Off-path SmartNIC for Accelerating Distributed Systems}
%\title{\Large \bf Everything you always wanted to know about transparent GPU remoting\\ but were afraid to ask}
\title{\Large \bf Characterizing Network Requirements for GPU API Remoting in AI Applications}

%\author{Xingda Wei, Rongxin Cheng, Yuhan Yang, Rong Chen, Haibo Chen}
%\affiliation{\institution
%{Shanghai Key Laboratory of Scalable Computing and Systems \\
%Institute of Parallel and Distributed Systems, Shanghai Jiao Tong University}
%}
%\email{Contact: rongchen@sjtu.edu.cn}

%\renewcommand{\shortauthors}{X. Wei et al.}

\iffalse
\author{Paper \#10}
\fi
% \iffalse
\setlength{\affilsep}{0.5em}
\author[1]{Tianxia Wang$^{*}$}
\author[2]{Zhuofu Chen\footnote{Tianxia Wang and Zhuofu Chen contribute equally to this work.
Work done while Zhuofu is intern at Institute of Parallel and Distributed Systems.
}}
\author[1]{Xingda Wei\thanks{Xingda Wei is the corresponding author (\url{wxdwfc@sjtu.edu.cn}).}}
\author[1]{Jinyu Gu}
\author[1]{Rong Chen}%\thanks{Rong Chen is the corresponding author (\url{rongchen@sjtu.edu.cn})}
\author[1] {Haibo Chen}
%\vspace*{-30pt}
\affil[1]{\vspace{-2.mm}Institute of Parallel and Distributed Systems, SEIEE, Shanghai Jiao Tong University\vspace{0.8mm}}
\affil[2]{Tongji University\vspace{-1.mm}}
% \fi
\date{}
\maketitle

\frenchspacing

% abstract
\begin{abstract}
  \noindent
  GPU remoting is a promising technique for supporting AI applications. 
  Networking plays a key role in enabling remoting.
  However, for efficient remoting, 
  the network requirements in terms of latency and bandwidth are unknown.
  In this paper, we take a GPU-centric approach 
  to derive the minimum latency and bandwidth requirements for GPU remoting, 
  while ensuring no (or little) performance degradation for AI applications. 
  Our study including theoretical model demonstrates that, with careful remoting design, 
  unmodified AI applications can run on the remoting setup using commodity networking hardware 
  without any overhead or even with better performance, with low network demands.    
\end{abstract}

% intro
\section{Introduction}
\label{sec:intro}

\noindent
GPU is a key pillar in speeding up artificial intelligence (AI) applications 
including DNN inference and training.
Most existing AI applications use the GPU locally---the 
applications running on the CPU invoke GPU APIs
(e.g., CUDA driver APIs) 
to execute computations directly on the GPU locally attached to the CPU. 
While local execution mode has been the norm for decades, 
recent works from both industry and academia suggest using GPUs remotely. 
In this approach (\emph{GPU API}\footnote{\footnotesize{We discuss 
GPU API remoting only in this paper, 
so we use the more general but shorter term ``API'' for GPU API.}} \emph{remoting}), 
CPUs invoke APIs
to a remote proxy that manages the GPUs through the network. 
Examples include disaggregated GPU datacenters~\cite{DBLP:conf/osdi/ShanHCZ18,DBLP:journals/corr/abs-2310-04648,DBLP:journals/corr/abs-2306-03622,DBLP:conf/hotos/HuWWSBZKZCXZFS23,guleria2019emf} 
and GPU virtualizations~\cite{DBLP:journals/concurrency/EilingBLM22,DBLP:conf/ieeehpcs/DuatoPSMQ10,DBLP:journals/corr/abs-2202-07848,DBLP:conf/sosp/NgD023,DBLP:journals/csur/HongSN17}.

The trends for using a remote proxy for API execution
are driven by the desire to improve the CPU/GPU resource efficiency of the traditional local execution mode, 
as well as the goal of enhancing the functionalities of (typically closed-source) GPUs (\textsection{\ref{sec:bg-app}}). 
For example, a local server nowadays has a fixed proportion of CPUs and GPUs (e.g., 16 vs. 1), 
but the application requirements for different resources are diverse (e.g., 12 CPUs vs. 1 GPU), 
especially in the cloud~\cite{DBLP:journals/corr/abs-2310-04648,DBLP:conf/osdi/ShanHCZ18}. 
Such a mismatch between the application requirements and the hardware setup may cause 
underutilized resources on the local server~\cite{DBLP:journals/corr/abs-2310-04648,DBLP:conf/osdi/ShanHCZ18}.
%This mismatched configuration may cause unused resources on the local server. 
API remoting allows CPUs and GPUs to be physically~\cite{DBLP:journals/corr/abs-2310-04648} 
or logically~\cite{DBLP:conf/ipps/FinglerZYJWR22,DBLP:journals/corr/abs-2306-03622} separated into different resource pools
(a.k.a., resource disaggregation).
This allows applications to select CPUs and GPUs independently in a more flexible manner, 
thereby improving the resource utilization.

Networking is a key enabler and performance impact factor when running AI applications with API remoting. 
This is because communication between the CPU and the proxy introduces 
non-trivial overhead (we called \emph{remoting overhead}) to the CPU-GPU communication.
Our initial evaluations show that 
existing open-source solution~\cite{DBLP:journals/concurrency/EilingBLM22} 
adds up to 27\,$\times$ overhead to AI applications.
While some closed-source solutions 
have shown improved remoting performance~\cite{DBLP:journals/corr/abs-2306-03622,DBLP:conf/ipps/FinglerZYJWR22},
whether their solutions are optimal and 
the detailed networking requirements for such improvements are still unknown. 
%which typically takes place over high-speed interconnects like PCIe with low latency and high bandwidth.  
%Our initial results show that even with fast networking like RDMA (2\,${\mu}$s and 200\,Gbps), 
%remoting can still add \red{xx--xx} overheads to AI applications.

Characterizing the network requirements 
for API remoting is crucial for both application and infrastructure developers. 
Considering the deployment of AI applications on 
a cloud platform backed by resource disaggregation~\cite{DBLP:journals/corr/abs-2310-04648,DBLP:conf/ipps/FinglerZYJWR22,guleria2019emf}. 
In this case, application developers aim to minimize the overhead introduced 
by disaggregation for better performance, 
and cloud vendors aim to minimize the network resources 
provisioned to the application for reducing the costs.
Unfortunately, and somewhat surprisingly, despite many efforts to realize and optimize API remoting, 
there has been little systematic characterization of the network requirements for efficient remoting, 
especially for AI applications. 
It is challenging for system designers and cloud vendors to decide on the right requirements 
when running applications in a remoting setting: 
different AI applications (and models) have varying network requirements, 
and different remoting technologies have different designs 
that can impact the performance under remoting.

In this paper, we conduct the first systematic study 
on deriving the \emph{minimum} latency and bandwidth requirements 
that the network must provide for an AI application under 
a performance-optimal remoting design. 
We define the minimal requirements as follows: 
for a given application and an acceptable overhead budget
(e.g., the added delay to the inference should be less than 5\%), 
the requirements ensure that the remoting overhead is within this budget.
This allows system designers to precisely balance the trade-off 
between the benefits of remoting and its overhead.

%\TODO{The relationship between the insight and the so-called GPU-centric approach is not clear. Besides, this insight seems somewhat trivial?}
A key insight behind our study is that 
\emph{the execution time of most AI applications is dominated by the GPU execution time}. 
Since the execution on the GPUs is triggered by the API,
the cumulative delay of remoting these APIs approximates remoting overhead. 
Based on this insight, we first adopt a \emph{GPU-centric} approach 
to derive a performance-optimal remoting design for AI applications, 
the goal is to minimize the delay of executing APIs on the GPU.
Then we adopt the same approach 
to formulate the remoting overhead as a function of network latency and bandwidth
by accumulating the theoretical delays of executing APIs on the GPU. 
Putting it all together,
we can systematically
derive the network requirements for running any AI applications
efficiently with API remoting.
In summary, our contributions are:

\stitle{A performance-optimal remoting design for AI applications. (\textsection{\ref{sec:design}})}
We summarize two principles to reduce all the possible delays of executing APIs 
on the GPU due to remoting: 
executing API asynchronously with outstanding requests (OR) 
and replicating GPU state with shadow descriptors (SR). 
Asynchronous execution is a canonical way to hide network delays. 
Unlike existing works that achieve async execution with batching~\cite{DBLP:journals/corr/abs-2306-03622,DBLP:conf/ipps/FinglerZYJWR22}, 
we show that async with OR can achieve a similar or better performance on modern networking (e.g., RDMA), 
and it does not rely on configuring the batch size to achieve an optimal performance. 
With SR, we can further transform an average of 79\% APIs from synchronous to asynchronous. 
We have incorporated these principles into an open-source remoting system~\cite{DBLP:journals/concurrency/EilingBLM22}. 
Applying both principles improves remoting performance by 3.8--100 \,$\times$ on microbenchmarks and 
up to 6 \,$\times$ on various AI applications.

\stitle{The first systematic formulation of the remote cost and remoting network requirements 
    for AI applications. 
    (\textsection{\ref{sec:model}})}
According to our system design, 
we are the first to formulate the remoting overhead as 
a function of the application's API calling pattern, network latency, and bandwidth. 
Based on this formulation, 
we can derive the network requirement for any AI applications that are bottlencked by GPU execution. 
We have also built a tool to analyzes the application pattern 
and automates the derivation of its network requirements.

\stitle{An end-to-end study on running AI applications with remoting. (\textsection{\ref{sec:app}})}
We use a combination of emulation and implementation on real hardware (i.e., RDMA), 
along with thorough evaluation on real-world AI applications, 
to demonstrate the efficacy of our optimization principles and veracity of our overhead formulation. 
We also study the impact of remoting on AI applications, 
including inference and training, using models ranging from large language models (GPT-2)
and traditional AI models like ResNet.
Our key findings---which matches our formulation---are that:  \\[-10pt]
\begin{itemize}[leftmargin=*,leftmargin=10pt,itemindent=0pt]
    \item Network latency is a key performance factor that 
    impacts AI applications. 
    Latency in the range of 5--20\,$\mu$s is sufficient 
    for most inference applications to benefit from API remoting with less than 5\% overhead.
    It is trivial to achieve so if the proxy is 
    co-located with the applications.    
    Meanwhile, 
    it is easy to achieve such latency in modern data centers network hardware.
    \\[-15pt]

    \item Many AI applications are not sensitive to the network bandwidth 
    between the CPU and GPU. For instance, even with a network latency of 10\,$\mu$s 
    and a provisioned bandwidth of 1\,Gbps, we only observe a small degradation of 2.5\% on GPT-2.
    \\[-15pt]

    \item The tolerance of overhead introduced by remoting is strongly 
    correlated with the applications' execution time on the GPU. 
    This correlation is derived from both our theoretical formulation and empirical results. 
    Since the execution time depends on the GPU, 
    the tolerance is also co-related to the GPU's performance.
%    Specifically, 
%    The longer the computation taken on the GPU, the lower the network requirement it needs.
    \\[-15pt]

    \item The tolerance for remoting overhead is 
    also strongly co-related to the application's features. 
    Training applications—--though runs longer than inference ones, 
    maybe more sensitive to network changes 
    due to its more frequent interactions between CPUs and GPUs.
    \\[-15pt]

    \item Finally, remoting may even improve the performance of AI applications 
    by overlapping of GPU and CPU execution time with async executions. 
    This is achievable on (not so new) commodity networking hardware (e.g., RDMA). 
\end{itemize}

Our study suggests a promising direction for adopting API remoting for AI applications. 
Under modern networking (200\,Gbps RDMA with 2--5\,$\mu$s RTT), 
we observe a maximum of 6\% overheads for various AI applications, 
with some applications even enjoy 1--14\% improvements.
Moreover, many of these applications do not require a powerful network for remote access. 
At one extreme, GPT-2 only experiences a 2\% performance degradation under 
a 10\,$\mu$s RTT and 1\,Gbps setup.
The main reason is that the execution time of AI applications on GPUs (10--100\,ms) is typically
significantly longer than the network delay in modern networking (1--100\,$\mu$s). 
Therefore, the incurred overhead can be amortized if there are no additional CPU-side software costs involved, 
such as unnecessary waiting of the completions of some GPUs' APIs. 

We will open-source our remoting system, 
benchmarks, and tools upon publication.

% bg
\section{Background: GPU API remoting}
\label{sec:motiv}

\subsection{Targeted system model}
\label{sec:bg-model}

\begin{figure}[!t]
    %    \vspace{-2mm}
        \begin{minipage}{1\linewidth}
        \hspace{-4mm}
        \centering    
        \includegraphics[scale=0.98]{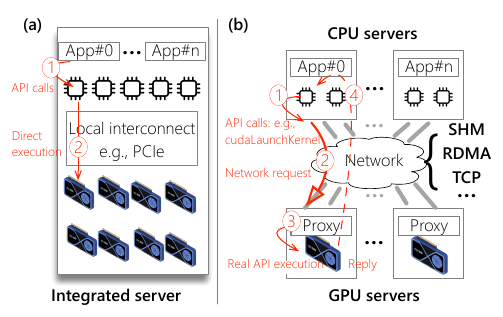}
        \end{minipage} \\[6pt]
        \begin{minipage}{1\linewidth}
        \caption{\small{%
        An overview of executing AI applications 
        with (a) and without (b) GPU remoting.
        }}
        \label{fig:setup}
        \end{minipage} \\[-15pt]
        \end{figure}

\vspace{-1.ex}
\stitle{GPU API remoting. }
AI applications commonly utilize GPU \emph{locally} ({\fig{fig:setup}} (a)):
The GPU is physically connected to the CPU running the application code, 
and the code running on the CPU interacts with the GPU with driver API calls (\ding{192})
through MMIO and DMA, 
both via local physical PCIe interconnects (\ding{193}).

Applications and systems may also use GPUs \emph{remotely}
for better resource utilizations and enhanced device capabilities~\cite{DBLP:conf/ipps/FinglerZYJWR22,DBLP:conf/ieeehpcs/DuatoPSMQ10,vmware-bit,DBLP:journals/tc/ShiCSL12,guleria2019emf}
(see \textsection{\ref{sec:bg-app}}).
%thanks to its compatibility with all existing GPU applications.
As shown in {\fig{fig:setup}} (b), 
the GPU is either physically located on a different machine 
or logically decoupled from the application (e.g., virtualization).
In this setup, applications interact with the GPU 
through a networked proxy that manages the GPU. 
Specifically, the CPU, which runs the application code, 
sends GPU driver API
calls (\ding{192}) to the proxy via the network (\ding{193}). 
The proxy then executes the commands on behalf of the application (\ding{194}) 
and returns the results back to the CPU (\ding{195}).

\stitle{Networking for GPU remoting.}
We study two networking primitives for GPU remoting in this paper, 
the fastest interconnects within a machine (SHM) and across machines (RDMA). \\[-10pt]
\begin{itemize}[leftmargin=*,leftmargin=10pt,itemindent=0pt]      
    \item \textbf{Shared memory communications (SHM).} 
    When the application and proxy are running on the same machine, 
    they can use shared memory for communication (e.g., via \texttt{mmap}). 
    Compared to other networking primitives, 
    systems that support remoting with SHM add minimal overhead to the above applications, 
    as local memory access latency and bandwidth are typically orders of magnitude higher than other remote accesses 
    (e.g., RDMA~\cite{farm} or CXL~\cite{DBLP:conf/sosp/ZhangMHLCDDJMW23}).

    \item \textbf{Remote direct memory access (RDMA). }
    RDMA is a low-latency and high-bandwidth networking feature widely deployed in modern data centers~\cite{tsai2017lite,DBLP:conf/sigcomm/GuoWDSYPL16}. 
    To send a networking request, the sender uses \texttt{post\_send}, 
    which internally issues a memory-mapped IO (MMIO) to notify the RDMA-capable NIC (RNIC) 
    to use direct memory access (DMA) for fetching the request payload and sending it. 
    Once the request has been sent to the receiver, 
    it can receive the request by reading its local memory via \texttt{poll\_completion}.
    \\[-3pt]
\end{itemize}

\subsection{Killer applications of transparent GPU remoting}
\label{sec:bg-app}

%\noindent
%There are two typical use cases of remoting: 

\begin{figure}[!t]
    \hspace{-1mm}
    \includegraphics[left, scale=1.2]{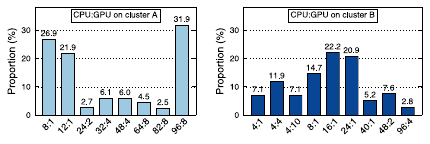} \\[1pt]
    \begin{minipage}{1\linewidth}
        \caption{\small{
        A re-plot of the CPU vs. GPU allocation configurations for applications 
        in two production clusters~\cite{DBLP:journals/corr/abs-2310-04648}.
        }}    
    \label{fig:data-motiv-resource}
\end{minipage} \\[-10pt]
\end{figure}

\begin{figure*}[!t]
    \hspace{-1mm}
    \includegraphics[left, scale=1.2]{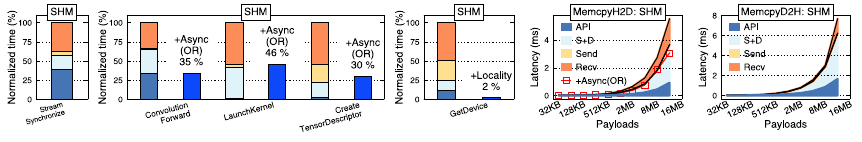} %\\[-1pt]
    \includegraphics[left, scale=1.2]{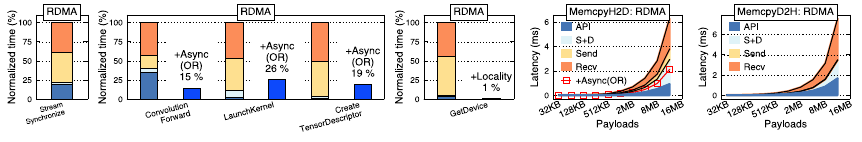} \\[-2pt]
    \begin{minipage}{1\linewidth}
            \caption{\small{
            Remoting overhead and optimization effects on A100
             with 
            (upper) SHM and (lower) RDMA, respectively. 
            We break down the remoting time as follows: 
            \textbf{API}---the execution time of the API without remoting, 
            \textbf{S+D}---serialization and deserialization of the arguments to the network buffer, 
            \textbf{Send}---posting the arguments to the network card and 
            \textbf{Recv}---the time for waiting for the proxy’s reply.   
            }}
    \label{fig:data-overhead-opt}
\end{minipage} \\[-5pt]
\end{figure*}
%\footnotetext{\footnotesize{Here is the text of the footnote.}}

\vspace{-1.ex}
\stitle{Improving resource utilizations via remoting-based disaggregation.}
GPU disaggregation
is a popular architecture aimed at enhancing CPU/GPU utilizations~\cite{DBLP:conf/ipps/FinglerZYJWR22,DBLP:journals/corr/abs-2310-04648,guleria2019emf,DBLP:conf/osdi/GaoNKCH0RS16,DBLP:conf/osdi/ShanHCZ18}.
Specifically, the CPUs and GPUs in a cluster are separated into CPU and GPU blades, 
where applications running on the CPUs utilize GPUs with remoting. 
The issue with traditional monolithic architecture is that the CPU vs. GPU ratio on a server is fixed, 
e.g., 128:8 in DGX-100~\cite{dgx100}. 
On the other hand, CPU vs. GPU configurations are highly dynamic in production clusters,
as shown in {\fig{fig:data-motiv-resource}}.
If, for example, an application needs a configuration of 128:7, one GPU is wasted on the assigned server. 
With GPU disaggregation, 
system administrators can assign CPUs and GPUs in a more flexible way
since the CPU can access any GPUs through the network.

\stitle{Enhancing device capabilities via remoting-based virtualization.} 
Modern GPUs typically delegate critical functionalities, 
such as resource management and scheduling, 
to the device hardware or (possibly closed-source, e.g., CUDA) driver. 
However, as the device hardware are challenging to evolve according to application demands, 
while closed-source driver is hard to add functionalities, 
recent works~\cite{DBLP:journals/corr/abs-2202-07848, DBLP:conf/sosp/NgD023} 
virtualize the device to enhance the device capabilities. 
Specifically, they use remoting to intercept GPU API calls to a proxy (similar to microkernel), 
where the proxy can implement customized capabilities.
For instance, 
Singularity~\cite{DBLP:journals/corr/abs-2202-07848} virtualized the GPU to support transparent 
device state checkpoint.
Paella~\cite{DBLP:conf/sosp/NgD023} virtualized the device to enable customized kernel scheduling.

% design
\section{Performance-optimal remoting design}
\label{sec:design}

\begin{figure*}[!t]
    %    \vspace{-2mm}
        \begin{minipage}{1\linewidth}
        \hspace{-4mm}
        \centering    
        \includegraphics[scale=1.1]{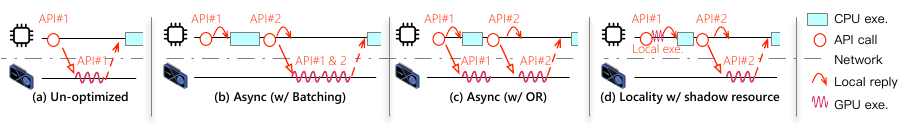}
        \end{minipage} \\[0pt]
        \begin{minipage}{1\linewidth}
        \caption{\small{%
        An illustration of the optimization space ((b)—(d)) for GPU API remoting, 
        comparing the unoptimized baseline (a). 
        }}
        \label{fig:opts}
        \end{minipage} \\[-5pt]
        \end{figure*}

\noindent
In this section, we analyze remoting overheads for dominating GPU APIs 
in AI applications (\textsection{\ref{sec:app}})
and present optimization guidelines.

% hardware-tab
% Save the current values
\newlength{\oldintextsep}
\newlength{\oldtextfloatsep}
\newlength{\oldbelowcaptionskip}
\setlength{\oldintextsep}{\intextsep}
\setlength{\oldtextfloatsep}{\textfloatsep}
\setlength{\oldbelowcaptionskip}{\belowcaptionskip}

% Set new values to reduce space
\setlength{\belowcaptionskip}{-8pt}
\setlength{\textfloatsep}{3pt}
\setlength{\intextsep}{3pt}

\begin{table}[!t]
    \vspace{2.2mm}
    \begin{minipage}{1\linewidth}
        \caption{\small{
            Meansurement clusters.
        }}
    \label{tab:cluster}
    \end{minipage} \\[-3pt]
    \hspace*{-5mm}
    \centering
    \small{
    \resizebox{.94\linewidth}{!}{
    \ra{1.1}
    \begin{tabular}{@{~}l@{~}l@{~}l@{~}}
    \toprule
    {\textbf{Name}}   & {\textbf{\#\,\,\,\,\,\,\,\,}} & {\textbf{Hardware Descriptions (Desc.)}}                                       \\ 
    \hline
    \textbf{V100}   & 2\,                 & 2$\times$  Intel Xeon CPU (12\,cores, 2.2\,GHz, 252\,GB DRAM)                               \\
             &                     & 1$\times$  ConnectX-6 IB RNIC (200\,Gbps)                                         \\
             &                     & 1$\times$  V100 (16\,GB HBM, SXM2)                                                                       \\
    \textbf{A100}   & 1\,                 & 2$\times$  Intel Golden CPU (56\,cores, 3.3\,GHz, 1\,TB DRAM)                               \\
             &                     & 2$\times$  ConnectX-6 IB RNIC (200\,Gbps)                                         \\
             &                     & 8$\times$  A100 (80\,GB HBM, SXM4)  \\
    \bottomrule      \\[-10pt]          
    \end{tabular} 
    }
    } 
    \end{table}  

    % Restore the original values
\setlength{\intextsep}{\oldintextsep}
\setlength{\textfloatsep}{\oldtextfloatsep}
\setlength{\belowcaptionskip}{\oldbelowcaptionskip}

\stitle{Evaluation setup and baseline implementations. }
We perform all our experiments on two local clusters listed in Table~\ref{tab:cluster}, 
named based on the GPUs they are equipped with.
Our baseline remoting framework is built on cricket~\cite{DBLP:journals/concurrency/EilingBLM22}, 
a state-of-the-art open-source GPU remoting framework. 
We have extended the it with three network backends 
for the application to communicate with the proxy:
(1) \emph{SHM} implements a local memory-based ring buffer 
for the proxy to communicate with the application, 
where the buffer is mapped to both processes' virtual address space via \texttt{mmap}.
(2) \emph{RDMA} adopts the messaging primitive from a 
state-of-the-art RDMA framework~\cite{drtm-h} that 
integrated with all known RDMA-aware optimizations. 
(3) \emph{Simulator} 
extends the SHM backend 
to emulate various network latency and bandwidth 
(\textsection{\ref{sec:design:emulation}} describes the emulation methodology in details).
We have also optimized the end-to-end software stack of the cricket, 
including replacing less efficient data structures (e.g., linked list) 
with more performant ones (e.g., hashtables). 

\stitle{Overhead analysis. }
We conducted a microbenchmark that iteratively executes a single GPU API. 
This benchmark analyzes the overhead introduced by remoting 
by comparing the average time with and without remoting 
on the A100 GPU\footnote{\footnotesize{Due to space limitations, 
we only present and discuss the results for A100 in \textsection{\ref{sec:design}}, 
the results on V100 are similar. }}
{\fig{fig:data-overhead-opt}} shows that executing the GPU APIs 
on the remote proxy without optimization ({\fig{fig:opts}} (a)) 
results in 3.6--58\,$\times$ and 3.9--102\,$\times$ overheads for different APIs 
under SHM and RDMA, respectively. 
The overheads can be broken down into two parts: 
the time for waiting for the remote proxy to send the reply (Recv + API), 
and the time introduced by software stacks for remoting, 
such as preparing the network request (serialization and deserialization, S+D) 
and posting the request to the NIC (Send). 
RDMA has a slightly higher Send overhead due to the relatively costly MMIO~\cite{drtm-h}, 
while SHM only requires a simple memory write.

\stitle{Basic optimization: async execution. }
A classic solution for hiding the aforementioned costs is 
remoting the APIs asynchronously (\textbf{async})~\cite{DBLP:journals/corr/abs-2306-03622,DBLP:conf/ipps/FinglerZYJWR22}, 
i.e., directly return without really executing the API (see {\fig{fig:opts}} (b), API\#1). 
Not immediately executing the API is correct 
because many GPU APIs are asynchronous by design, i.e., 
the return value is irrelevant to the execution of the API.
%This means that their results won't affect the execution of the CPU. 
For example, the semantic of the \texttt{launchkernel}\footnote{\footnotesize{Since 
we only evaluated NVIDIA GPUs in this paper, 
we may omit the \texttt{CUDA} prefix in API calls without losing generality.}}
API is that the kernel will eventually be launched on the GPU.

\begin{figure}[!t]
    %    \vspace{-2mm}
        \begin{minipage}{1\linewidth}
        \hspace{-2mm}
%        \centering    
        \includegraphics[scale=1.2]{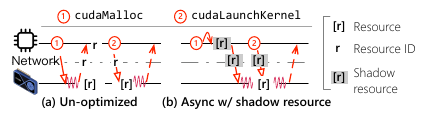}
        \end{minipage} \\[1pt]
        \begin{minipage}{1\linewidth}
        \caption{\small{%
        An illustration of how shadow resource enables a synchronous API (\texttt{CreateTensorDescriptor}, \ding{192}) 
        to become asynchronous, while remaining compatible with APIs that depend on it (e.g., 
        \texttt{ConvolutionForward} \ding{193}).
        }}
        \label{fig:opt-shadow}
        \end{minipage} \\[-4pt]
        \end{figure}

\stitle{Challenges for async execution.}
Two challenges arise when considering applying async optimization.
(1) How to timely send the API to the proxy while balancing the overhead of issuing the network request?
Existing solutions adopt batching~\cite{DBLP:journals/corr/abs-2306-03622,DBLP:conf/ipps/FinglerZYJWR22} 
to reduce the overhead of network requests ({\fig{fig:opts}} (b)), 
but the execution of the API on the GPU, for example API\#1, 
may be delayed. 
Since most AI applications are dominated by GPU execution, 
delaying the sending of API requests (e.g., launching a kernel later) 
with batching may even degrade the overall performance of application 
(2) How to make more APIs async while preserving their semantics? 
Not all GPU APIs are async (see Table~\ref{tab:app-data}). 
For example, \texttt{cudaMalloc} allocates a memory buffer on the GPU and 
returns the pointer. 
It is non-async because the CPU cannot obtain the pointer without executing it on the GPU.
Existing solutions simply execute them synchronously~\cite{DBLP:journals/corr/abs-2306-03622}, 
which cannot hide the remoting overhead. 

To this end, 
we summarize the following two principles for 
designing performance-optimal remoting: 

% app-table
\begin{table*}[!t]
  \vspace{2.2mm}
  \begin{minipage}{1\linewidth}
      \caption{\small{
      A characterization of the percentage of the types of APIs 
      executed by different AI applications and their execution time (evaluated on RDMA). 
      ``\textbf{\uline{+SR}}'' means the shadow resource.         
      Higher \textbf{Async} and \textbf{Local} numbers implies better remoting performance.
      }}
  \label{tab:app-data}
  \end{minipage} \\[-1pt]
%    \hspace*{-5mm}
  \centering
  \small{
  \resizebox{.99\textwidth}{!}{
  \ra{1.2}
  \begin{tabular}{l|cccccccc|cccccccc}
      \toprule
      & \multicolumn{1}{l}{} & \multicolumn{1}{l}{}              & \multicolumn{3}{l}{\textbf{Application: Inference (\# API)}}                        & \multicolumn{1}{l}{}              & \multicolumn{1}{l}{} & \multicolumn{1}{l|}{} & \multicolumn{1}{l}{} & \multicolumn{1}{l}{}              & \multicolumn{3}{l}{\textbf{Application: Inference (API time)}}                      & \multicolumn{1}{l}{}              & \multicolumn{1}{l}{} & \multicolumn{1}{l}{} \\ \cline{4-6} \cline{12-14}
             & \multicolumn{1}{l}{} & \multicolumn{1}{l}{}              & \multicolumn{1}{l}{} & \multicolumn{1}{l}{}              & \multicolumn{1}{l}{} & \multicolumn{1}{l}{}              & \multicolumn{1}{l}{} & \multicolumn{1}{l|}{} & \multicolumn{1}{l}{} & \multicolumn{1}{l}{}              & \multicolumn{1}{l}{} & \multicolumn{1}{l}{}              & \multicolumn{1}{l}{} & \multicolumn{1}{l}{}              & \multicolumn{1}{l}{} & \multicolumn{1}{l}{} \\[-8pt]
             & \textbf{ResNET}      & \multicolumn{1}{c|}{\textbf{+SR}} & \textbf{SD}          & \multicolumn{1}{c|}{\textbf{+SR}} & \textbf{BERT}        & \multicolumn{1}{c|}{\textbf{+SR}} & \textbf{GPT-2}       & \textbf{+SR}          & \textbf{ResNET}      & \multicolumn{1}{c|}{\textbf{+SR}} & \textbf{SD}          & \multicolumn{1}{c|}{\textbf{+SR}} & \textbf{BERT}        & \multicolumn{1}{c|}{\textbf{+SR}} & \textbf{GPT-2}       & \textbf{+SR}         \\ \hline
             \textbf{Async} & 414                                  & \multicolumn{1}{l|}{\highlight{534}}          & 149,053                          & \multicolumn{1}{l|}{\highlight{169,193}}       & 467                               & \multicolumn{1}{l|}{467}          & 6,104                               & 6,104                              & 0.51\,ms                             & \multicolumn{1}{l|}{0.58\,ms}      & 0.30 s                          & \multicolumn{1}{l|}{0.30\,s}       & 1.05\,ms                           & \multicolumn{1}{l|}{1.14\,ms}      & 9.28\,ms                            & 14.49\,ms                         \\
             \textbf{Local} & 0                                    & \multicolumn{1}{l|}{\highlight{937}}          & 0                                & \multicolumn{1}{l|}{\highlight{583,968}}       & 0                                 & \multicolumn{1}{l|}{\highlight{2,407}}         & 0                                  & \highlight{37,634}                                & 0.00 ms                             & \multicolumn{1}{l|}{0.03 ms}      & 0.00 s                          & \multicolumn{1}{l|}{0.02\,s}       & 0.01\,ms                           & \multicolumn{1}{l|}{0.07\,ms}      & 0.01\,ms                            & 1.05\,ms                          \\
             \textbf{Sync}  & 1,061                                & \multicolumn{1}{l|}{4}                        & 607,831                          & \multicolumn{1}{l|}{3,723}         & 2,436                              & \multicolumn{1}{l|}{29}            & 38,145                              & 511                               & 3.70\,ms                             & \multicolumn{1}{l|}{0.05\,ms}      & 5.92\,s                          & \multicolumn{1}{l|}{5.57 s}       & 9.18 ms                           & \multicolumn{1}{l|}{0.45\,ms}      & 110.80\,ms                          & 6.74\,ms                          \\ \hline
             \textbf{Total} & 1,475                                & \multicolumn{1}{l|}{1,475}                    & 756,884                          & \multicolumn{1}{l|}{756,884}       & 2,903                              & \multicolumn{1}{l|}{2,903}         & 44,249                              & 44,249                             & 4.21\,ms                             & \multicolumn{1}{l|}{\highlight{0.66\,ms}}      & 6.22 s                          & \multicolumn{1}{l|}{\highlight{5.89\,s}}       & 10.23\,ms                          & \multicolumn{1}{l|}{\highlight{1.66\,ms}}      & 120.07\,ms                          & \highlight{22.29\,ms}                         \\
      \bottomrule
\end{tabular}
  }
  } 
  \end{table*}

\etitle{Principle \#1:\,async with outstanding request (OR). }            
Our key observation is that modern networking like RDMA 
allows us to async without batching, 
because the overhead of posting a network request (Send)
is consistent low on modern networking hardware.
For example, a single MMIO in RDMA only requires 400--800 cycles, 
which may be negligible compared to the time spent waiting 
for the CPU to issue the next batches of APIs. 
Other fast networking techniques such as DPDK also has a similar time~\cite{lim2014mica}.
Therefore, we can directly send the network request without waiting for it to complete, 
i.e., making the request outstanding. 
As shown in {\fig{fig:opts}} (c), with OR, API\#1 
is executed earlier at the proxy than async with batching (b). 

The challenge of OR is to ensure that 
the order in which requests arrive at the proxy matches the order in which they are sent. 
This can be trivially achieved, 
for example, all requests sent to the same RDMA connection 
is sent (and received) through the network in a FIFO order~\cite{ib}.

\etitle{Principle \#2:\,apply shadow descriptor (SR) and locality. }
This principle systematically converts many sync APIs to async ones.
The observation is that for sync APIs that follow a resource-related pattern, 
there is no need to execute it synchronously. 
Specifically, for a resource-related API, 
the CPU typically calls it to allocate 
a type of resource (e.g., buffer, tensor) ({\fig{fig:opt-shadow}} (a)).
Since the resource is only used by GPU, we can overlap its creation 
with the CPU execution using async. 
We convert it to async APIs as follows. 
We logically replicate each resource 
with a corresponding shadow component on the CPU-side (SR). 
As shown in {\fig{fig:opt-shadow}} (b), 
to allocate a buffer on the GPU, 
the CPU can simply create a shadow buffer 
and then forward the create API with OR asynchronously without waiting for it.

One issue introduced by SR is that future APIs may reference the (shadow) resource,
e.g., \texttt{LaunchKernel} needs to compute on the created buffer.
Since the proxy is unaware of the shadow resource,
directly using its reference for the other API results in correctness issues.
We fix this by forwarding the shadow resource reference to the proxy in the OR request.
Hence, the proxy can establish a mapping between the shadow and the real ID, 
so it can alter the IDs timely for correctness.

Replicating the resources on the CPU-side further allows us 
to completely avoid the remoting overhead by executing some APIs locally 
(\emph{locality} in {\fig{fig:opts}} (d)). 
Specifically, there is no need to send read-only requests 
querying the resource state to the remote proxy.
For example, by replicating which GPU device is used by each CPU thread in its SR, 
we can execute the \texttt{GetDevice} locally.

\stitle{Results: improving API execution. }
{\fig{fig:data-overhead-opt}} presents the optimization results to 
each API's execution time. 
For sync APIs like \texttt{StreamSynchronize} and copying memory from the GPU to the CPU (\texttt{MemcpyD2H}), 
there is little optimization space on the system’s perspective. 
For async APIs, OR can improve the end-to-end API execution time by 26--75\%
for APIs with static payloads (e.g., \texttt{LaunchKernel}), 
and 32--68\% when copying memory from the CPU to the device for payloads between 32\,KB and 16\,MB. 
Note that several APIs are not async by default (e.g., \texttt{CreateTensorDescriptor}); 
SR makes them so.
Finally, locality enables a 95\% improvements for localizable APIs (i.e., \texttt{GetDevice}).

\begin{figure}[!t]
    \hspace{-2mm}
    \includegraphics[left, scale=1.15]{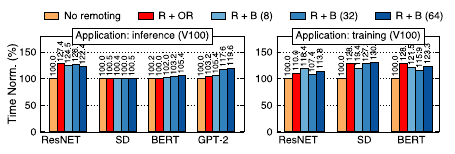} \\[1pt]
    \begin{minipage}{1\linewidth}
            \caption{\small{
            A comparison of async execution with batching and OR. 
            The results are normalized to non-remoting performance.
            }}
    \label{fig:data:batching}
\end{minipage} \\[-10pt]
\end{figure}

\vspace{1pt}
Improvements to the single API may not necessarily result in faster application performance, 
e.g., if it is rarely called.
Next, we use application benchmarks to show 
the effectiveness of the summarized principles in end-to-end application execution time. 
The detailed setups can be found in \textsection{\ref{sec:app}}.

\stitle{Results: batching vs. OR. }
{\fig{fig:data:batching}} compares OR with batching. 
The reported numbers are normalized to the non-remoting performance.
We can draw two observations from the figure.
First, OR has similar a similar performance as batching:
it is up to 15\%\,$\times$ faster (and 19\%\,$\times$ slower than the optimal batch size)
compared to batching for various batch configurations.
Inference is less sensitive to batching size 
because its execution pattern is simpler. 
During inference, the CPU only spawns kernels for the forward pass. 
On the other hand, during training, 
the CPU must alternate between forward, backward, and parameter update kernels.
Second, it is hard to decide the batch size for training: 
a smaller batch cannot amortize the overhead of invoking RDMA request
 (e.g., B=8 in ResNET training), 
 while a larger batch may delay the execution of GPU kernels 
 and result even in performance overhead (e.g., SD training with B=64). 
OR achieves comparable performance without the need to configure the batch size.

\stitle{Results: effects of SR and locality. }
Table~\ref{tab:app-data} summarizes the API patterns 
and their corresponding execution time for inference applications
with and without SR and locality optimizations.
%We run the applications with Pytorch, a standard framework for AI.
To collect the data, we profile the types of APIs called 
of a specific application and record their execution time. 
Note that an application may call an API multiple times 
and we will record them faithfully. 
This is important because the overall execution time 
is related to the cumulative API times (see \textsection{\ref{sec:model}}). 

We can see that without SR, 72--84\%
APIs are sync for various models inference tasks. 
When SR is enabled, 88--98\% of them are converted to local ones, 
and 0--11\% are converted to async ones, respectively. 
An interesting observation is that modern AI frameworks 
have many local APIs. 
For example, 
PyTorch frequently calls \texttt{GetDevice} along in a \texttt{DeviceGuard} 
to ensure that each kernel is launched on the correct GPU stream. 
By default, \texttt{GetDevice} is sync, 
so remoting has has a huge remoting overhead to the applications of PyTorch.
Nevertheless, 
By replicating the device setting with SR, we can transform it into a local API.
The benefits of transformations are obvious: 
the overall remoting overhead (API time) has been reduced from 
6--84\% for various models.
Next we will analyze its contributions to the overall performance. 

\begin{figure}[!t]
    \hspace{-2mm}
    \includegraphics[left, scale=1.15]{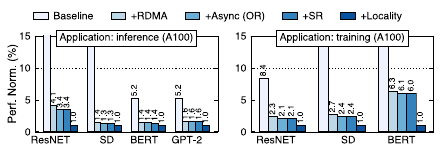} \\[3pt]
    \begin{minipage}{1\linewidth}
            \caption{\small{
        A factor analysis of the optimizations 
        and comparison with existing solutions.
            }}
    \label{fig:data:factor}
\end{minipage} \\[-10pt]
\end{figure}

\stitle{Factor analysis. }
Finally, {\fig{fig:data:factor}} shows a factor analysis 
of the optimization of our TCP-based backend~\cite{DBLP:journals/concurrency/EilingBLM22}. 
We cumulatively add various optimizations to it. 
All the reported numbers are normalized to the fastest baseline 
(when all the optimizations have been enabled).
We use the RDMA backend and omit the results on SHM since the trend is similar.
We can observe that all optimizations contribute to improving overall performance, 
although the extent of improvement may vary. 
First, RDMA reduces the execution time 
by 69--93\% and 72--88\% for inference and training applications,
respectively.  
The improvements come from replacing
the slow kernel TCP/IP stack with kernel-bypassing and hardware accelerated RDMA stack. 
It is not surprising to see such an improvements 
given the orders of magnitude faster network performance of RDMA.  
Second, locality plays a crucial role in remote communication. 
This is primarily because Pytorch's execution mode heavily relies on resource query APIs, 
which have minimal overhead in a local setting. Without locality, 
all of these APIs have to be treated as sync APIs 
so they would significantly slow down the application performance.

% model
\section{Cost model for GPU remoting}
\label{sec:model}

\noindent
Based on the previous performance-optimal remoting design, 
this section derives a cost model to theoretically analyze the overhead introduced by remoting.

%\vspace{-1.ex}
\stitle{GPU-centric remoting cost analysis.}
%To simplify the derivation, 
%We take a \emph{GPU-centric} approach, 
%meaning the application's execution time is 
The key insight of our derivation is that 
most AI application's execution time is 
dominated by the GPU's execution.
Since the computation on the GPU is triggered by the API, 
the accumulation of delayed API arrival on the GPU 
approximates the overhead introduced by remoting.
To further simplify derivation, 
we assume CPU calls APIs sequentially, 
so we can simply accumulate the delays. 
Considering concurrency would require 
reasoning the overlapping between communication and computation 
which is more complicated. 
Fortunately, our evaluated AI applications 
rarely calls GPU API concurrently\footnote{\footnotesize{
    In training applications, Pytorch's data loaders~\cite{pytorch-dataloader} may 
    concurrently 
    load the data and then copy the data to the GPU (with remoting).    
    Nevertheless, the copy can still be approximated as a sequential 
    transfer since they go through the same link.
}}. 

\begin{figure}[!t]
    %    \vspace{-2mm}
        \begin{minipage}{1\linewidth}
        \hspace{-2mm}
%        \centering    
        \includegraphics[scale=1.2]{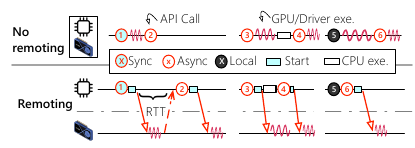}
        \end{minipage} \\[5pt]
        \begin{minipage}{1\linewidth}
        \caption{\small{%
        An illustration of how we derive the remoting cost model 
        for different types of APIs.
        }}
        \label{fig:model}
        \end{minipage} %\\[-20pt]
        \end{figure}

\stitle{Cost model ($Cost(APP)$ ).}
We use a bottom-up approach 
by first deriving the cost of each individual API, 
then profile all the APIs called by an application, 
and finally accumulate them to get the remoting cost. 
We define the remoting cost as the time differences
between the local execution and the remoting execution.
Since the costs of async and sync APIs are different, 
we derive them separately in {\eqa{eq:cost}}: 

\setlength{\abovedisplayskip}{4pt} % Adjust the space accordingly
\setlength{\abovedisplayshortskip}{4pt} % Adjust the space for "short" equations
\begin{equation}
\label{eq:cost}
\begin{aligned}
    C_{async}(api)  = Start(api) + \frac{RTT}{2} + \frac{Payload(api)}{Bandwidth}  \\
    C_{sync}(api)   = Start(api) + RTT  + \frac{Payload(api)}{Bandwidth}
\end{aligned}
\end{equation} \\[-5pt]

\noindent
$Start$ is the software overhead to send a network request, 
which includes the time to post the request to the network card
 (e.g., one MMIO in RDMA)
and the time of serialization and deserialization (S+D) the arguments to the 
message buffer. 
The $Start$ overhead is almost constant for each API.
$RTT$ is the network round-trip time, i.e., 
the the for a packet from the application-side
to the proxy-side. 
$Payload$ is the arguments size of the API,
and $B$ is the network bandwidth.
Note that for sync APIs, the $Payload$ also includes the response data.
For most APIs, $\frac{Payload(api)}{B}$ is negligible.
This is because their payload is small, e.g., 64\,B. 
Such a small payload can be sent inline with the network packet~\cite{DBLP:conf/usenix/KaliaKA16}.

Two things need to be noted in our derivation. 
(1) We calculate the network delay of a sync API as $RTT$ instead of $\frac{RTT}{2}$, 
even though the API is only delayed for $\frac{RTT}{2}$ when executing on the GPU. 
As shown in {\fig{fig:model}}, \ding{192} is only delayed for $\frac{RTT}{2}$ 
when executing on the GPU.
The additional $\frac{RTT}{2}$ is from the delay of the response, 
which delays the execution of the next API.
As in {\fig{fig:model}}, (\ding{193}) is delayed 
due to the response of its previous API (\ding{192}).
(2) We don't consider the API execution time for the sync APIs 
since it is the same regardless of the local or remote execution.

\vspace{1ex}
When considering the async and locality optimizations (\textsection{\ref{sec:design}}), 
an interesting phenomenon is that
remoting accelerate the arrival of severals APIs on the GPU,
which indirectly accelerate the execution of them.
{\fig{fig:model}} illustrates this.
For async APIs, take API (\ding{194}) as an example:
it accelerates the execution of its next API (\ding{195}) compared to the local execution,
because the application's CPU execution can be overlapped with the GPU API execution.
Besides, 
for API with locality optimization (e.g., {\ding{196}}), 
it also will accelerate the next API 
if the execution on the shadow resource is faster than on the real resource.
We formulate these accelerations as follows (\eqa{eq:opt}): 

\begin{equation}
\label{eq:opt}
\begin{aligned}
    E_{local}(api) = Time(api) - Time_{local}(api)  \\
    E_{async}(api) = Time(api)
\end{aligned}
\end{equation} \\[3pt]

\begin{figure*}[!t]
    \hspace{-1mm}
    \includegraphics[left, scale=1.2]{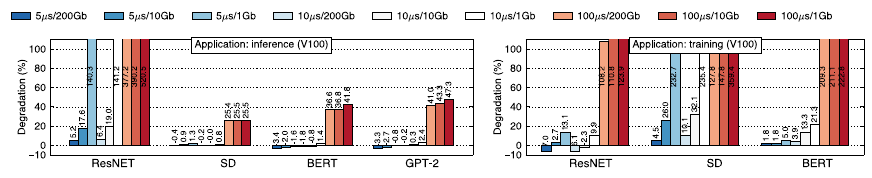} %\\[-1pt]
    \includegraphics[left, scale=1.2]{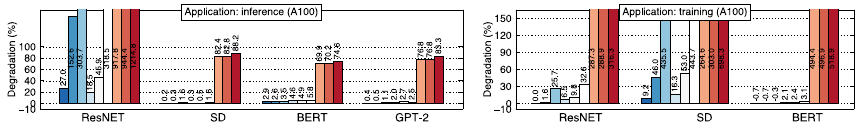} %\\[-1pt]
    \begin{minipage}{1\linewidth}
            \caption{\small{
            Comparison of application performance based on network latency/bandwidth configurations 
            on V100 (upper) and A100 (lower).
            }}
    \label{fig:requirements}
\end{minipage} \\[-5pt]
\end{figure*}

Put it all together,
given a network configuration ($RTT$ and $Bandwidth$ and $Send_{network}$) , 
we can formulate the performance cost introduced 
by remoting for a given application ($APP$) 
by accumulating \eqa{eq:cost} and \ref{eq:opt}: \\[-10pt]
{
\begin{multline}
    \label{eq:overall}
    Cost(APP) = 
    \sum_{api}^{APP_{AsyncAPIs}} (C_{async}(api) - E_{async}(api)) \\
    + \sum_{api}^{App_{SyncAPIs}} C_{sync}(api) - \sum_{api}^{App_{LocalAPIs}} E_{local}(api)
\end{multline}
}

\stitle{Deriving network requirements. }
Given an application $APP$, 
to constrain the remoting overhead 
to be within a given threshold ($\epsilon$),
we can simply deduce the network requirements 
by searching a network configuration ($RTT$ and $Bandwidth$)
such that $Cost(APP) \leq \epsilon$. 
For most AI applications, 
the $\epsilon$ can be estimated by multiplying an overhead budget 
(e.g., 5\%) with the expected execution time 
(e.g., forward time and iteration time in local execution 
for inference and training applications, respectively).
\textsection{\ref{sec:eval-app-hardware}} 
further validates the accuracy of our model under real hardware setups. 

% app
\section{Network requirements for AI applications}
\label{sec:app}

\noindent
Based on our system designs and cost model,
we now characterize 
the network requirements for running AI applications remotely.

% app-desc-tab
\setlength{\belowcaptionskip}{25pt}
\setlength{\textfloatsep}{3pt}
\setlength{\intextsep}{5pt}

\begin{table}[!t]
    \vspace{-8.0mm}
    \begin{minipage}{1\linewidth}
        \caption{\small{
        Applications evaluated. 
        All the models and code are pulled from HuggingFace~\cite{huggingface}. 
        }}
    \label{tab:app-desc}
    \end{minipage} \\[-1pt]
%    \hspace*{-5mm}
    \centering
    \small{
    \resizebox{.99\linewidth}{!}{
    \ra{1.2}
    \begin{tabular}{l|ccc}
        \toprule
        & \textbf{Inference} & \textbf{Training} & \textbf{Desc.}       \\ \hline
\textbf{ResNET}                &      \cmark                            &   \cmark                                    & Vision model                  \\
\textbf{SD}                    &      \cmark                            &   \cmark                                    & Stable diffusion                   \\
\textbf{BERT}                  &      \cmark                            &   \cmark                                    & Language model                     \\
\textbf{GPT-2}                 &      \cmark                            &   -                                         & Large language model               \\
    \bottomrule
\end{tabular}  
    }
    } 
\end{table}

% Restore the original values
\setlength{\intextsep}{\oldintextsep}
\setlength{\textfloatsep}{\oldtextfloatsep}
\setlength{\belowcaptionskip}{\oldbelowcaptionskip}

\stitle{Evaluated applications. }
We run all the applications using PyTorch with 
models, inference, and training code from HuggingFace~\cite{huggingface}.
Table~\ref{tab:app-desc} summarizes the applications used in our evaluations, 
which involves representative models used by AI applications. 
For example, ResNET represents vision models with CNNs, 
while GPT-2 and Bert represents transformers with and without auto-regressive 
execution, respectively. 
The parameters scale are 11M, 1.1B, 110M and 1.5B
for ResNET, SD, Bert and GPT-2, respectively.
Since the training of GPT-2 is similar to Bert, 
we omit it for simplicity. 

\stitle{Reported metrics.}
For inference, we reported (and compared) the time it took to 
finish the forward pass of a batch, including the time the CPU reads the results from the (remote) GPU.  
For training, we reported the time to finish an iteration.  
We pre-warm both of the applications to rule out the interference from initialization. 

\subsection{Characterization network requirements}
\label{sec:design:emulation}

\noindent
To evaluate a broad range of network configurations, 
we first build a network emulator based 
on the SHM-backend to emulate various network configurations, 
i.e., RTT and bandwidth.

%\vspace{-1.ex}
\stitle{Metholodgy. }
Similar to previous work~\cite{DBLP:conf/osdi/GaoNKCH0RS16}, 
we inject artificial delays based on the configuration of the network
to emulate its configuration.
Specifically, 
the application assigns an expected arrival time to each remoting API request, 
allowing the proxy to check and wait until that time to process it. 
Since the application and proxy run on the same machine under SHM, 
their time is synchronized, 
so the proxy can simply check the time based on its current clock.
At the proxy-side, 
we implement a priority queue that processes requests according to the 
timestamps. 
Our emulation considered the queuing between requests sent by the same application
 by regulating the delay based on the current inflight requests. 
Yet, our injected latency does not account high network tail latency, 
like a priori work~\cite{DBLP:conf/osdi/GaoNKCH0RS16}. 
We have verified that the software overhead of the emulation is negligible 
compared to the network delays we added. 

\stitle{Results. }
We begin by evaluating the performance degradation of 
remoting compared to non-remoting 
across different network latency and bandwidth configurations. 
{\fig{fig:requirements}} plots the results. 
Unlike previous work~\cite{DBLP:conf/osdi/GaoNKCH0RS16},
our evaluated latencies are pure-hardware---the software 
overhead including message serialization and deserialization has been 
included in the application execution.
These software overheads are irrelevant to the network  
so we don't consider them as a network latency. 

% bandwidth-tab
\setlength{\belowcaptionskip}{-8pt}
\setlength{\textfloatsep}{3pt}
\setlength{\intextsep}{3pt}

\begin{table}[!t]
    \vspace{2.2mm}
    \begin{minipage}{1\linewidth}
        \caption{\small{
        Bandwidth requirements for different applications 
        on V100 (left) and A100 (right). 
        We use share memory to calculate the network transmission data size in 
        the local execution setup. 
        }}
    \label{tab:app-bandwidth}
    \end{minipage} \\[-1pt]
    \hspace*{-5mm}
    \centering
    \small{
    \resizebox{.99\linewidth}{!}{
    \ra{1.2}
    \begin{tabular}{l|cccccccc}
        \toprule
        & \multicolumn{2}{c}{\textbf{ResNET}} & \multicolumn{2}{c}{\textbf{SD}} & \multicolumn{2}{c}{\textbf{Bert}} & \multicolumn{2}{c}{\textbf{GPT-2}} \\ \hline
\textbf{Inference (MB/s)} & 253     & 279.4       & 0.8            &   1.2           & 0.6              & 0.9           & 0.25               & 0.4            \\
\textbf{Training (MB/s)}  & 12.3          & 24.6               & 220.4            & 390.8         & 0.02           & 0.03              & -             & -           \\
        \bottomrule
\end{tabular}    
    }
    } 
    \end{table}  

    % Restore the original values
\setlength{\intextsep}{\oldintextsep}
\setlength{\textfloatsep}{\oldtextfloatsep}
\setlength{\belowcaptionskip}{\oldbelowcaptionskip}

For inference, 
we made two observations from {\fig{fig:requirements}}.
First, our applications can be divided into two categories: 
high requirements and low requirements.
For example, 
on V100, ResNET requires a 5\,$\mu$s and 200\,Gbps configuration to achieve 
a 5\% performance penalty, 
while SD, Bert, and GPT only require 10\,$\mu$s and lower bandwidth (e.g., 1\,Gbps). 
On A100, the requirement of ResNET is also 5\,$\mu$s and 200\,Gbps, 
while the requirements of SD, Bert, and GPT are 10us and 1\,Gbps. 
The key reason for difference between models is their inference time:
the model with a longer execution (e.g., 4.3\,ms in ResNET inference vs. 580\,ms in GPT inference in V100) 
typically can tolerate a larger network delay.
Second, most applications are more sensitive to latency rather than bandwidth, except for ResNET.
Under 10\,$\mu$s setup, the degradations of SD, Bert, and GPT are similar 
when configuring the network from 200\,Gbps to 1\,Gbps, 
while ResNET suffers a 135--277\% execution time increase.
This is due to two reasons. (1) As we have discussed before, 
ResNET has a shorter execution time, 
making it more sensitive to the additional latency 
introduced by longer data transfer time caused by constrained bandwidth.
(2) ResNET has a higher bandwidth requirements than others: 
we have profiled the bandwidth of different applications in a non-remoting setting 
in Table~\ref{tab:app-bandwidth}. 
We can see that ResNET requires a high bandwidth, 
e.g., 253\,MB/s vs. 0.6\,MB/s (in Bert) than the others.
This is partly because its inference time is shorter and 
its data transfer for each batch element is larger (image vs. text). 

When it comes to training, 
the impact of network is more evident to the application performance. 
For example, when the latency of network increases from 5\,$\mu$s--100\,$\mu$, 
the iteration time suffers a degradations of 1.8--209.3\% for Bert, respectively. 
Though the GPU execution time is longer in training, 
it also needs extensive interactions with the CPU for the forward, loss calculation, 
backward and parameter update passes, so its trends are not directly comparable to the inference. 
ResNET is less sensitive to network downgrades compared to Bert 
since its execution time is longer on GPU. 
An exception is SD: though it has a longer execution time, 
it also suffers 400\% degradation when configuring with a 100\,$\mu$s latency.

Finally, A100 needs a higher network requirements than V100. 
For example, the degradations of Bert inferences 
increase from -3.4-- -1.8\% to 2.9--4.6\% when configuring 
the network from 5\,$\mu$s to 100\,$\mu$s with a fixed 200\,Gbps bandwidth. 
This is also because the applications' execution time 
is shorter on A100 because of its more powerful capabilities. 
Therefore, we need a faster network to hide the remoting overhead.

\begin{figure}[!t]
    \hspace{-2mm}
    \includegraphics[left, scale=1.15]{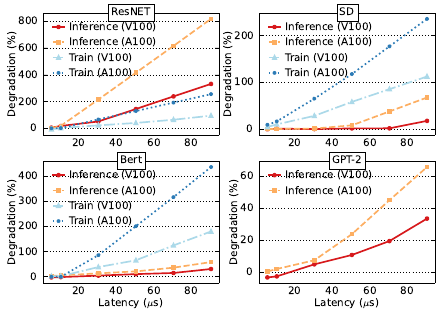} \\[3pt]
    \begin{minipage}{1\linewidth}
            \caption{\small{
            Impact of network latency on the results of {\fig{fig:requirements}} 
            with a fixed 200\,Gbps bandwidth configuration.                
            }}
    \label{fig:data:sensitiveness}
\end{minipage} \\[-5pt]
\end{figure}

% app-data-table

\begin{table*}[!t]
  \vspace{2.2mm}
  \begin{minipage}{1\linewidth}
      \caption{\small{
      Comparison of inference adn training performance with \textbf{\uline{R}}emoting on real hardware, 
      i.e., V100 (top) and A100 (bottom). 
      ``\textbf{\uline{+opt}}'' adopts all the optimizations described in \textsection{\ref{sec:design}}.
       We evaluated two \textbf{\uline{B}}atch sizes: 
       1 and the number that saturates the GPU. 
       ``\textbf{\uline{+theo}}'' states for the theoretical 
       cost calculated using \eqa{eq:overall}.  
       Note that a batch size of one is sufficient for SD to saturate the GPU.  
       The blue shadow below the numbers indicates the ratio 
       compared to the slowest baseline.        
      }}
  \label{tab:inference-data}
  \end{minipage} \\[2pt]
  \hspace*{-5mm}
  \centering
  \small{
  \resizebox{.94\textwidth}{!}{
  \ra{1.2}
  % v100-tab
  \begin{tabular}{l|ccccccc|ccccccc}
    \toprule
    & \textbf{}                                                       & \multicolumn{1}{l}{\textbf{}}                                    & \multicolumn{3}{c}{\textbf{V100 inference time (ms)}}                                                                                                                                        & \multicolumn{1}{l}{}                                            & \multicolumn{1}{l|}{}                                            & \multicolumn{3}{c}{\textbf{V100 training time (ms)}}                                                                                                                                            \\ \cline{4-6} \cline{9-11} 
    & \textbf{\begin{tabular}[c]{@{}c@{}}ResNET\\ (B=1)\end{tabular}} & \textbf{\begin{tabular}[c]{@{}c@{}}ResNET\\ (B=64)\end{tabular}} & \textbf{\begin{tabular}[c]{@{}c@{}}SD\\ (B=1)\end{tabular}} & \textbf{\begin{tabular}[c]{@{}c@{}}Bert\\ (B=1)\end{tabular}} & \textbf{\begin{tabular}[c]{@{}c@{}}Bert\\ (B=64)\end{tabular}} & \textbf{\begin{tabular}[c]{@{}c@{}}GPT-2\\  (B=1)\end{tabular}} & \textbf{\begin{tabular}[c]{@{}c@{}}GPT-2\\ (B=512)\end{tabular}} & \textbf{\begin{tabular}[c]{@{}c@{}}ResNET\\ (B=64)\end{tabular}} & \textbf{\begin{tabular}[c]{@{}c@{}}SD\\ (B=1)\end{tabular}} & \textbf{\begin{tabular}[c]{@{}c@{}}Bert\\ (B=64)\end{tabular}} \\ \hline
    \textbf{Local}          & \PartiallyFilledCellWidth{appcolor}{A}{1}{0.360}4.3                              & \PartiallyFilledCellWidth{appcolor}{A}{1}{0.484}5.8                               & \PartiallyFilledCellWidth{appcolor}{A}{1}{0.980}8118.3                        & \PartiallyFilledCellWidth{appcolor}{A}{1}{0.542}17.8                              & \PartiallyFilledCellWidth{appcolor}{A}{1}{0.881}50.5                               & \PartiallyFilledCellWidth{appcolor}{A}{1}{0.429}185.5                               & \PartiallyFilledCellWidth{appcolor}{A}{1}{0.804}596.9                                   & \PartiallyFilledCellWidth{appcolor}{A}{1}{0.738}65.8                    & \PartiallyFilledCellWidth{appcolor}{A}{1}{0.586}776.9                   & \PartiallyFilledCellWidth{appcolor}{A}{1}{0.318}55.8                   \\
    \textbf{+ R (SHM)}      & \PartiallyFilledCellWidth{appcolor}{A}{1}{0.668}8.0                              & \PartiallyFilledCellWidth{appcolor}{A}{1}{0.680}8.2                               & \PartiallyFilledCellWidth{appcolor}{A}{1}{0.988}8180.3                        & \PartiallyFilledCellWidth{appcolor}{A}{1}{0.675}22.2                              & \PartiallyFilledCellWidth{appcolor}{A}{1}{0.871}50.0                               & \PartiallyFilledCellWidth{appcolor}{A}{1}{0.715}309.2                               & \PartiallyFilledCellWidth{appcolor}{A}{1}{0.859}637.6                                   & \PartiallyFilledCellWidth{appcolor}{A}{1}{0.818}73.0                    & \PartiallyFilledCellWidth{appcolor}{A}{1}{0.791}1049.9                  & \PartiallyFilledCellWidth{appcolor}{A}{1}{0.615}105.5                  \\
    \textbf{+ R (SHM+opt)}  & \PartiallyFilledCellWidth{appcolor}{A}{1}{0.257}3.1                              & \PartiallyFilledCellWidth{appcolor}{A}{1}{0.507}6.1                               & \PartiallyFilledCellWidth{appcolor}{A}{1}{0.978}8103.4                        & \PartiallyFilledCellWidth{appcolor}{A}{1}{0.444}14.6                              & \PartiallyFilledCellWidth{appcolor}{A}{1}{0.853}48.9                               & \PartiallyFilledCellWidth{appcolor}{A}{1}{0.324}140.1                               & \PartiallyFilledCellWidth{appcolor}{A}{1}{0.763}566.8                                   & \PartiallyFilledCellWidth{appcolor}{A}{1}{0.732}65.3                    & \PartiallyFilledCellWidth{appcolor}{A}{1}{0.593}786.6                   & \PartiallyFilledCellWidth{appcolor}{A}{1}{0.326}52.7                   \\
    \textbf{+ R (RDMA)}     & \PartiallyFilledCellWidth{appcolor}{A}{1}{1.000}12.0                             & \PartiallyFilledCellWidth{appcolor}{A}{1}{1.000}12.0                              & \PartiallyFilledCellWidth{appcolor}{A}{1}{1.000}8282.0                        & \PartiallyFilledCellWidth{appcolor}{A}{1}{1.000}32.9                              & \PartiallyFilledCellWidth{appcolor}{A}{1}{1.000}57.3                               & \PartiallyFilledCellWidth{appcolor}{A}{1}{1.000}432.4                               & \PartiallyFilledCellWidth{appcolor}{A}{1}{1.000}742.5                                   & \PartiallyFilledCellWidth{appcolor}{A}{1}{1.000}89.2                    & \PartiallyFilledCellWidth{appcolor}{A}{1}{1.000}1326.8                  & \PartiallyFilledCellWidth{appcolor}{A}{1}{1.000}172.7                  \\
    \textbf{+ R (RDMA+opt)} & \PartiallyFilledCellWidth{appcolor}{A}{1}{0.305}3.7                              & \PartiallyFilledCellWidth{appcolor}{A}{1}{0.514}6.2                               & \PartiallyFilledCellWidth{appcolor}{A}{1}{0.979}8107.4                        & \PartiallyFilledCellWidth{appcolor}{A}{1}{0.477}15.7                              & \PartiallyFilledCellWidth{appcolor}{A}{1}{0.847}48.5                               & \PartiallyFilledCellWidth{appcolor}{A}{1}{0.367}158.8                               & \PartiallyFilledCellWidth{appcolor}{A}{1}{0.766}568.6                                   & \PartiallyFilledCellWidth{appcolor}{A}{1}{0.764}68.2                    & \PartiallyFilledCellWidth{appcolor}{A}{1}{0.598}793.5                   & \PartiallyFilledCellWidth{appcolor}{A}{1}{0.344}62.0                   \\
    \textbf{+ R (RDMA+theo)} & \PartiallyFilledCellWidth{appcolor}{A}{1}{0.334}4.0                             & \PartiallyFilledCellWidth{appcolor}{A}{1}{0.475}5.7                               & \PartiallyFilledCellWidth{appcolor}{A}{1}{0.936}7752.4                        & \PartiallyFilledCellWidth{appcolor}{A}{1}{0.502}16.5                              & \PartiallyFilledCellWidth{appcolor}{A}{1}{0.860}49.3                               & \PartiallyFilledCellWidth{appcolor}{A}{1}{0.377}163.1                               & \PartiallyFilledCellWidth{appcolor}{A}{1}{0.795}590.1                                   & \PartiallyFilledCellWidth{appcolor}{A}{1}{0.737}65.8                    & \PartiallyFilledCellWidth{appcolor}{A}{1}{0.672}891.0                   & \PartiallyFilledCellWidth{appcolor}{A}{1}{0.347}59.0                   \\
    \bottomrule
\end{tabular}
  }
  \\[10pt]
  \resizebox{.94\textwidth}{!}{
  \ra{1.2}
  % a100-tab
  \begin{tabular}{l|ccccccc|ccccccc}
    \toprule
    & \textbf{}                                                       & \multicolumn{1}{l}{\textbf{}}                                    & \multicolumn{3}{c}{\textbf{A100 inference time (ms)}}                                                                                                                                        & \multicolumn{1}{l}{}                                            & \multicolumn{1}{l|}{}                                            & \multicolumn{3}{c}{\textbf{A100 training time (ms)}}                                                                                                                                            \\ \cline{4-6} \cline{9-11} 
    & \textbf{\begin{tabular}[c]{@{}c@{}}ResNET\\ (B=1)\end{tabular}} & \textbf{\begin{tabular}[c]{@{}c@{}}ResNET\\ (B=64)\end{tabular}} & \textbf{\begin{tabular}[c]{@{}c@{}}SD\\ (B=1)\end{tabular}} & \textbf{\begin{tabular}[c]{@{}c@{}}Bert\\ (B=1)\end{tabular}} & \textbf{\begin{tabular}[c]{@{}c@{}}Bert\\ (B=64)\end{tabular}} & \textbf{\begin{tabular}[c]{@{}c@{}}GPT-2\\  (B=1)\end{tabular}} & \textbf{\begin{tabular}[c]{@{}c@{}}GPT-2\\ (B=512)\end{tabular}} & \textbf{\begin{tabular}[c]{@{}c@{}}ResNET\\ (B=64)\end{tabular}} & \textbf{\begin{tabular}[c]{@{}c@{}}SD\\ (B=1)\end{tabular}} & \textbf{\begin{tabular}[c]{@{}c@{}}Bert\\ (B=64)\end{tabular}} \\ \hline
    \textbf{Local}           & \PartiallyFilledCellWidth{appcolor}{A}{1}{0.222}2.7        & \PartiallyFilledCellWidth{appcolor}{A}{1}{0.218}2.7      & \PartiallyFilledCellWidth{appcolor}{A}{1}{0.718}5093.1       & \PartiallyFilledCellWidth{appcolor}{A}{1}{0.311}8.6         & \PartiallyFilledCellWidth{appcolor}{A}{1}{0.677}34.6       & \PartiallyFilledCellWidth{appcolor}{A}{1}{0.227}83.7         & \PartiallyFilledCellWidth{appcolor}{A}{1}{0.616}385.4        & \PartiallyFilledCellWidth{appcolor}{A}{1}{0.430}30.7      & \PartiallyFilledCellWidth{appcolor}{A}{1}{0.372}414.4      & \PartiallyFilledCellWidth{appcolor}{A}{1}{0.160}28.6     \\
    \textbf{+ R (SHM)}       & \PartiallyFilledCellWidth{appcolor}{A}{1}{0.430}5.2        & \PartiallyFilledCellWidth{appcolor}{A}{1}{0.451}5.6      & \PartiallyFilledCellWidth{appcolor}{A}{1}{0.730}5175.5       & \PartiallyFilledCellWidth{appcolor}{A}{1}{0.501}13.8        & \PartiallyFilledCellWidth{appcolor}{A}{1}{0.697}35.6       & \PartiallyFilledCellWidth{appcolor}{A}{1}{0.445}163.8        & \PartiallyFilledCellWidth{appcolor}{A}{1}{0.680}425.0        & \PartiallyFilledCellWidth{appcolor}{A}{1}{0.606}43.3      & \PartiallyFilledCellWidth{appcolor}{A}{1}{0.614}683.1      & \PartiallyFilledCellWidth{appcolor}{A}{1}{0.313}55.7     \\
    \textbf{+ R (SHM-opt)}   & \PartiallyFilledCellWidth{appcolor}{A}{1}{0.127}1.5        & \PartiallyFilledCellWidth{appcolor}{A}{1}{0.211}2.6      & \PartiallyFilledCellWidth{appcolor}{A}{1}{0.719}5098.5       & \PartiallyFilledCellWidth{appcolor}{A}{1}{0.248}6.8         & \PartiallyFilledCellWidth{appcolor}{A}{1}{0.685}35.0       & \PartiallyFilledCellWidth{appcolor}{A}{1}{0.178}65.5         & \PartiallyFilledCellWidth{appcolor}{A}{1}{0.623}389.8        & \PartiallyFilledCellWidth{appcolor}{A}{1}{0.421}30.1      & \PartiallyFilledCellWidth{appcolor}{A}{1}{0.387}430.5      & \PartiallyFilledCellWidth{appcolor}{A}{1}{0.154}27.5     \\
    \textbf{+ R (RDMA)}      & \PartiallyFilledCellWidth{appcolor}{A}{1}{1.000}12.1       & \PartiallyFilledCellWidth{appcolor}{A}{1}{1.000}12.5     & \PartiallyFilledCellWidth{appcolor}{A}{1}{1.000}7092.3       & \PartiallyFilledCellWidth{appcolor}{A}{1}{1.000}27.6        & \PartiallyFilledCellWidth{appcolor}{A}{1}{1.000}51.1       & \PartiallyFilledCellWidth{appcolor}{A}{1}{1.000}368.3        & \PartiallyFilledCellWidth{appcolor}{A}{1}{1.000}625.3        & \PartiallyFilledCellWidth{appcolor}{A}{1}{1.000}71.4      & \PartiallyFilledCellWidth{appcolor}{A}{1}{1.000}1113.3     & \PartiallyFilledCellWidth{appcolor}{A}{1}{1.000}178.3    \\
    \textbf{+ R (RDMA-opt)}  & \PartiallyFilledCellWidth{appcolor}{A}{1}{0.164}2.0        & \PartiallyFilledCellWidth{appcolor}{A}{1}{0.218}2.7      & \PartiallyFilledCellWidth{appcolor}{A}{1}{0.719}5100.8       & \PartiallyFilledCellWidth{appcolor}{A}{1}{0.265}7.3         & \PartiallyFilledCellWidth{appcolor}{A}{1}{0.695}35.5       & \PartiallyFilledCellWidth{appcolor}{A}{1}{0.194}71.3         & \PartiallyFilledCellWidth{appcolor}{A}{1}{0.629}393.4        & \PartiallyFilledCellWidth{appcolor}{A}{1}{0.438}31.3      & \PartiallyFilledCellWidth{appcolor}{A}{1}{0.391}435.1      & \PartiallyFilledCellWidth{appcolor}{A}{1}{0.159}28.3     \\
    \textbf{+ R (RDMA-theo)} & \PartiallyFilledCellWidth{appcolor}{A}{1}{0.259}3.1        & \PartiallyFilledCellWidth{appcolor}{A}{1}{0.253}3.2      & \PartiallyFilledCellWidth{appcolor}{A}{1}{0.704}4993.5       & \PartiallyFilledCellWidth{appcolor}{A}{1}{0.334}9.2         & \PartiallyFilledCellWidth{appcolor}{A}{1}{0.717}36.6       & \PartiallyFilledCellWidth{appcolor}{A}{1}{0.256}94.1         & \PartiallyFilledCellWidth{appcolor}{A}{1}{0.656}410.2        & \PartiallyFilledCellWidth{appcolor}{A}{1}{0.476}34.0      & \PartiallyFilledCellWidth{appcolor}{A}{1}{0.467}520.0      & \PartiallyFilledCellWidth{appcolor}{A}{1}{0.204}36.4     \\
    \bottomrule
\end{tabular}
  }    
  } 
  \end{table*}  

\stitle{Sensitive analysis. }
According to {\fig{fig:requirements}},
most applications are more sensitive to network latency.
We next evaluate their sensitiveness to RTT in {\fig{fig:data:sensitiveness}}, 
which plots the performance degradation 
under increased network RTT with a fixed 200\,Gbps bandwidth.

The first observation is that the degradation of all applications 
increases almost linearly with network RTT. 
This is not surprising: based on our formulation in {\eqa{eq:cost}}, 
the cost is proportional to RTT. 
Second, 
different applications have different slopes, 
which explains why some applications are more sensitive to network latency than others.
For example, ResNET has a much steeper slope compared to SD in inference. 
Typically, a steeper slope indicates a higher requirement for the networking, 
because the increase of network latency would result in a larger overhead. 
For ResNET, increasing the RTT from 5 to 100\,$\mu$s increases the overhead by 353\%,
while SD only has 25\%. 

The factors that affect the slope of an application can be derived from 
{\eqa{eq:overall}} and {\eqa{eq:cost}}.
A clear factor is the RTT of the network. 
Except the RTT, 
the execution time of the application also plays an important role. 
Specifically, 
the degradation is linear with applications execution time multiples 
by the degradation ratio, 
so the slope is inversely proportional to the application time. 
Therefore, a longer execution time of an AI application 
makes it more tolerable to network delays introduced by remoting.

We should note that it is challenging to compare different 
applications and models, 
since the number of sync/async APIs also affect the slopes of the applications. 
Therefore, even the training applications may have a longer execution time, 
it slopes may be also steeper (see 
the training of Bert and SD as an example).

Finally, an interesting observation is that the slope also indicates 
the difference of running GPUs. 
We can see that 
the more powerful the GPU, the steeper the slope,
which matches our observation that 
a faster GPU requires a faster network configuration.

\begin{figure}[!t]
    \hspace{-2mm}
    \includegraphics[left, scale=1.15]{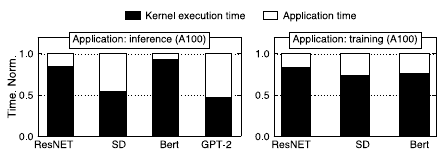} \\[3pt]
    \begin{minipage}{1\linewidth}
            \caption{\small{
        A profile of the GPU kernel execution time and the application execution time.        
            }}
    \label{fig:data:coleration}
\end{minipage} \\[-5pt]
\end{figure}

\stitle{Unstanding the results. }
The key derivation and results are based on the observation 
that the execution time of the application is dominated by the GPU execution time. 
As a result, the overhead introduced by remoting will not significantly affect the CPU-side execution. 
{\fig{fig:data:coleration}} profiles the application execution time 
of the end-to-end application time and the computation taken on GPU (GPU kernel time). 
We can see that for all our inference and training applications, the observation hold. 

\subsection{When requirements meet real hardware}
\label{sec:eval-app-hardware}

% cluster-network-table
\setlength{\belowcaptionskip}{-8pt}
\setlength{\textfloatsep}{3pt}
\setlength{\intextsep}{3pt}

\begin{table}[!t]
    \vspace{1.5mm}
    \begin{minipage}{1\linewidth}
        \caption{\small{
            Network configurations 
            of the measurement clusters.
            The latencies and bandwidths are measured with 
            RDMAPerf~\cite{rdma-perf} (
            \texttt{ib\_send\_lat} and \texttt{ib\_send_bw}). 
        }}
    \label{tab:cluster-network}
    \end{minipage} \\[1pt]
    \hspace*{-5mm}
    \centering
    \small{
    \resizebox{.86\linewidth}{!}{
    \ra{1.1}
    \begin{tabular}{lcc}
    \toprule
        & \textbf{Latency} & \textbf{RDMA Bandwidth (Measured)} \\
    \hline        
   \textbf{V100} &   2.6\,$\mu$s               &     200\,Gbps (180\,Gbps)               \\
   \textbf{A100} &   4.5\,$\mu$s               &     200\,Gbps (180\,Gbps)               \\
   \bottomrule
   \end{tabular}        
    }
    } 
    \end{table}  

    % Restore the original values
\setlength{\intextsep}{\oldintextsep}
\setlength{\textfloatsep}{\oldtextfloatsep}
\setlength{\belowcaptionskip}{\oldbelowcaptionskip}

\noindent
Table~\ref{tab:inference-data} 
summarizes the overall performance of inference 
and training on real V100 and A100 hardware clusters.
%Overall, when all optimizations described in the \textsection{\ref{sec:design}} are enabled, 
On SHM that represents the optimal network configuration (100\,ns latency and 600\,GBps),
remoting (SHM+opt) is consistently not slower (or faster)
than local execution on both V100 and A100 clusters. 
On RDMA (A100),
the remoting performance (RDMA+opt) is close (less than 2--5\%) 
to local execution for various applications (inference training), 
models, and batch sizes (whether saturated or not). 
For some applications (e.g., ResNET), RDMA+opt is even 25\% faster. 
Their network configurations of RDMA are listed in Table~\ref{tab:cluster-network}.
By comparing it with our emulation results, 
we clearly see that the configurations in both clusters satisfies 
the requirements of inference and training applications.
Therefore, the performance of degradation is small. 

There is no surprise in obtaining such empirical results:
They are guided by the theoretical
and emulation results described in 
\textsection{\ref{sec:model}} and \textsection{\ref{sec:design:emulation}}, respectively.
Take ResNET as an example. 
On V100, its inference time for batch 64 is 3.7\,ms, which is close 
to the emulation results in {\fig{fig:requirements}}.
Moreover, the theoretical results (RDMA+theo) calculated by \eqa{eq:overall}
are also close to our empirical results. 
We calculate the result by substituting the network configuration of RDMA
with the profiled API execution and serialization/deserialization time in \eqa{eq:cost}.
This confirm the validity of our theoretical model.

Note that the results on real hardware may 
deviate from our theoretical (and emulation) model. 
This is due to the fact that we are unable to model the queuing delays 
and resource contentions, 
as well as the fact that the profile of several constants
(e.g., $Start$ and $Time(api)$ in \eqa{eq:cost}) 
may have fluctuations. 

\subsection{Feasibility and takeaways}
\label{sec:takeaways}

\vspace{-1.ex}
\stitle{Feasibility of the networking requirements. }
It is trivial to see that for SHM the aforementioned network 
requirements (mainly RTT) for various applications, such as 5--20\,$\mu$s, 
 can be achieved. 
Moreover, we argue that such requirements
can be achieved in both local clusters (see the above section) 
and real-world data centers. 
From our derivation, 
faster execution on the GPU leads to tighter requirements (lower latency) on the network.  
The fastest execution is ResNET inference, with requirements of 
5\,$\mu$s and 200\,Gbps for V100 and A100, respectively.

The network latency of modern networking has long surpassed the requirement of 5--20\,$\mu$s. 
For example, modern RDMA-capable NICs can achieve latencies as low as 600\,ns,
with a bandwidth of 400\,Gbps~\cite{connect7}.
Nevertheless,
in real data centers, 
we need to further consider the latencies introduced by network topology, 
including (1) data transmission within and between racks, 
and (2) cut-through switches. 
The detailed cost depends on the scale and topology of the data centers. 
Previous studies~\cite{DBLP:conf/osdi/GaoNKCH0RS16,cachecloud} have shown that 
the cost has continuously decreased and has approached the physical limits. 
According to Gao et al.~\cite{DBLP:conf/osdi/GaoNKCH0RS16}, 
for common datacenter setups 
(e.g., 40\,m inter-rack and 4\,m intra-rack, 
with a propagation speed of 5\,ns/m, 
6 and 2 switch traversals for inter- and intra-rack), 
the RTTs for intra-rack and inter-rack are 1.38\,$\mu$s and 3.14\,$\mu$s, respectively, 
which are lower than our requirements.

In general, AI applications can tolerate higher GPU remoting latency 
compared to remoting other part of the hardware in a local server. 
For example, 
disaggregated memory (DM) is a popular architecture such that 
the CPU access the remote memory using network~\cite{DBLP:conf/osdi/GaoNKCH0RS16,DBLP:conf/usenix/TsaiSZ20}. 
A priori work~\cite{DBLP:conf/osdi/GaoNKCH0RS16}
has shown that DM require a latency as shorter as 3\,$\mu$ 
for common applications, which are much shorter than our requirements.
Compared to remoting memory, remoting GPUs requires a lower network configuration
because
(1) most APIs are asynchronous in their semantics, 
allowing us to hide the latency without modifying the applications, 
and (2) the execution time on GPU is much longer than the network time 
(ms vs. $\mu$s), 
leaving room for tolerating execution delays.

\stitle{Takeaways. }
We conclude our study by summarizing the takeaways.
Since we have discussed 
the detailed network requirements on our hardware platform before, 
we summarize the general findings irrelevant to a specific hardware configuration:
 \\[-10pt]
\begin{itemize}[leftmargin=*,leftmargin=10pt,itemindent=0pt]       
    \item For AI applications that require infrequent data transfer, 
    the degradations under remoting are linearly related to the RTT.  
    Therefore, the degradation introduced 
    by remoting depends on the slopes of the RTT-degradation function.
    Applications with flatter slopes require lower network requirements.    

    \item The slope of the RTT-degradation graph is sensitive to various factors: 
    it is negatively correlated with the execution of applications, 
    and it also depends on the pattern of applications, 
    such as training and inference.
    The application pattern is important 
    because it determines how many APIs are async and how many are sync, 
    whose contribution to the slope is different ({\eqa{eq:cost}}).  
    Due to this diversity, we cannot always conduct an apples-to-apples comparison 
    between the tolerance of remoting overhead for different applications (and models).
    Nevertheless, we can still use our formulation ({\eqa{eq:overall}}) 
    to decide the requirements in a case-by-case way. 

    \item Finally, we should mention that our analysis 
    is based on the assumption that the computation is GPU-centric.
    While this assumption holds true for most AI applications, 
    it is important to note that other GPU applications may have different features, 
    so our results may not be applicable.     

\end{itemize} 

% dislim
\section{Discussion and limitation}
\label{sec:dislim} 

\vspace{-1.ex}
\stitle{Limitation.}
First, we only consider AI applications that can fit within a single GPU in this study. 
Modern AI applications such as LLM inference~\cite{alpaserve,pope2023efficiently} 
and training~\cite{narayanan2021efficient} may 
deployed across multiple GPUs. 
When running these applications with remoting, 
it is necessary to intercept and coordinate communications between GPUs, 
which is more complicated.
We plan to study the impact of GPU communications with remoting in the future. 

Second, our approach primarily focuses on applications with a GPU-centric computing model, 
where CPUs only issue commands (APIs) to the GPU and wait for the GPU's responses. 
While this pattern is common in many AI applications, 
such as inference and training, 
there are also applications that have more complex interaction patterns. 
For example, in reinforcement learning~\cite{ray}, 
when GPUs are performing computations, CPUs may concurrently run simulations, 
so only considering API delays for characterizing the remoting overhead may be insufficient.
%We plan to study these applications in our future work.

Finally, our study lacks empirical results on the queuing delays 
that commonly occur when different applications share the same cluster.
Modeling such delays are much harder 
since they depend on the applications' co-location patterns 
and how the underlying network do the congestion control~\cite{DBLP:conf/osdi/GaoNKCH0RS16}. 
Nevertheless, our formulation described in \textsection{\ref{sec:model}} 
still holds under such delays (with increased RTTs). 
%We plan to study these delays with real-world traces in our future work.

\stitle{Future trends.}
Our study mainly focus on A100 and V100 GPU and ConnectX-6 RDMA networking. 
Future GPUs are still becoming faster, 
while the network latency may be further reduced by advanced interconnects 
such as NVLink and CXL. 
Since some of our results are related to detailed hardware performance, 
they may change due to hardware upgrades.
Nevertheless, the principles, methods, and tools used in our study are general, 
and we believe they can be applied to different hardware configurations.

% related
\section{Related work}
\label{sec:dislim}

\vspace{-1.ex}
\stitle{GPU API remoting and virtualization. }
GPU API remoting has been extensively studied in the HPC and system communities, 
e.g., for GPU virtualization and disaggregation~\cite{DBLP:conf/hpdc/BecchiSGPRC12,
    DBLP:conf/cluster/GimenoPMBQ15,
    DBLP:conf/ieeehpcs/DuatoPSMQ10,
    DBLP:conf/europar/DuatoIMPQS09,
    DBLP:conf/icpp/DuatoPSMQ11,
    DBLP:conf/europar/GiuntaMAC10,
    DBLP:conf/eurosys/GuptaGSKTTR09,
    DBLP:conf/usenix/GuptaSTTR11,
    DBLP:journals/tpds/ZhangYQYG14,
    DBLP:journals/corr/abs-2202-07848,
    DBLP:conf/sosp/NgD023,
    DBLP:conf/ipps/FinglerZYJWR22,
    DBLP:journals/corr/abs-2306-03622,
    DBLP:journals/corr/abs-2310-04648,
    DBLP:journals/corr/abs-2306-03622,
    DBLP:conf/ipps/FinglerZYJWR22,
    duato2010modeling}.
We build upon these works and are the first to 
systematically summarize all the principles to accelerate API remoting, 
as well as systematically characterize 
the overhead introduced by remoting with all optimizations enabled.
DGSF~\cite{DBLP:conf/ipps/FinglerZYJWR22} and FaaSwap~\cite{DBLP:journals/corr/abs-2306-03622} 
both leverage batching for asynchronous API execution. 
We show that batching is not necessary with OR. 
Moreover, we propose using SR to transform more APIs from sync to async for a better performance.
Duato et al.~\cite{duato2010modeling} and DxPU~\cite{DBLP:journals/corr/abs-2310-04648} 
characterized the overhead of remoting under fast networking without optimizations, 
as summarized in our study. 
These optimizations are crucial to remoting performance and are more complex to formulate,
which we systematically formulate them in our study. 
Moreover, we provide a more in-depth study of 
how AI applications perform under remoting with different network configurations.

\stitle{Resource disaggregation.}
Remoting is widely used in resource disaggregated datacenters~\cite{DBLP:conf/osdi/WangMLLRNBNKX20,
racehashing,polardb,DBLP:conf/osdi/RuanSAB20,DBLP:conf/usenix/TsaiSZ20,DBLP:conf/osdi/ShanHCZ18,DBLP:conf/hotos/HuWWSBZKZCXZFS23}, 
which is a popular paradigm nowadays. 
Gao et al.~\cite{DBLP:conf/osdi/GaoNKCH0RS16} characterized the network requirements for disaggregated memory.
Since disaggregated GPU is also an important component in 
resource disaggregated datacenters~\cite{DBLP:journals/corr/abs-2310-04648,DBLP:journals/corr/abs-2306-03622,DBLP:conf/ipps/FinglerZYJWR22,guleria2019emf},
we characterize the requirements for disaggregating GPUs with API remoting.
In contrast to a purely empirical study conducted in Gao’s work,
we further utilize the workload patterns of AI applications,
i.e., the GPU-centric computing model
to formulate the network requirements for AI applications.

\stitle{Systems powered by remoting. }
Besides resource disaggregation,  
systems can also use API remoting to support  GPU sharing, 
fault tolerance and application migration, just to name a few~\cite{DBLP:journals/corr/abs-2202-07848,DBLP:conf/sosp/NgD023,DBLP:conf/ipps/FinglerZYJWR22,DBLP:journals/corr/abs-2306-03622}. 
Though these functionalities can be supported by user frameworks, 
e.g., Pytorch support migration with TorchElastic~\cite{torch-elastic}, 
systems based on API remoting have the benefits of transparency. 
Specifically, systems can support these functionalities for unmodified applications, 
which is particular important when deploying them in the cloud. 
Our study demonstrated that 
using remoting with SHM implementation will not result in evident overhead
for applications.

% concl

\section{Conclusion}
\label{sec:concl}

\noindent
This paper summarizes the principles for efficient GPU API remoting, 
and provides a comprehensive study on the network requirements for API remoting for AI applications. 
The key takeaway is that for AI applications that are dominated by GPU execution, 
we can systematically derive their network requirements given an overhead budget.
With extensive evaluations, 
conducted with both emulation and real hardware, 
demonstrate the effectiveness of our derivation. 
They also reveal the patterns behind how different AI applications can tolerate the overhead of remoting.
Though the takeaways and observations of our study depend 
on specific applications, 
the methodology, performance model and tools
are general and can be applied to other applications and scenarios.
We believe that the results and tools from our study can be beneficial 
for future research and deployment in AI remoting.

\balance

\small{
\bibliographystyle{acm}
\bibliography{gpu}

\begin{thebibliography}{10}

\bibitem{ib}
{\sc Association., I.~T.}
\newblock Infiniband architecture specification.
\newblock \burl{https://cw.infinibandta.org/document/dl/7859}, 2022.

\bibitem{DBLP:conf/hpdc/BecchiSGPRC12}
{\sc Becchi, M., Sajjapongse, K., Graves, I., Procter, A.~M., Ravi, V.~T., and
  Chakradhar, S.~T.}
\newblock A virtual memory based runtime to support multi-tenancy in clusters
  with gpus.
\newblock In {\em The 21st International Symposium on High-Performance Parallel
  and Distributed Computing, HPDC'12, Delft, Netherlands - June 18 - 22,
  2012\/} (2012), D.~H.~J. Epema, T.~Kielmann, and M.~Ripeanu, Eds., {ACM},
  pp.~97--108.

\bibitem{polardb}
{\sc Cao, W., Zhang, Y., Yang, X., Li, F., Wang, S., Hu, Q., Cheng, X., Chen,
  Z., Liu, Z., Fang, J., Wang, B., Wang, Y., Sun, H., Yang, Z., Cheng, Z.,
  Chen, S., Wu, J., Hu, W., Zhao, J., Gao, Y., Cai, S., Zhang, Y., and Tong,
  J.}
\newblock Polardb serverless: {A} cloud native database for disaggregated data
  centers.
\newblock In {\em {SIGMOD} '21: International Conference on Management of Data,
  Virtual Event, China, June 20-25, 2021\/} (2021), G.~Li, Z.~Li, S.~Idreos,
  and D.~Srivastava, Eds., {ACM}, pp.~2477--2489.

\bibitem{DBLP:conf/cluster/GimenoPMBQ15}
{\sc Castell{\'{o}}, A., Pe{\~{n}}a, A.~J., Mayo, R., Balaji, P., and
  Quintana{-}Ort{\'{\i}}, E.~S.}
\newblock Exploring the suitability of remote {GPGPU} virtualization for the
  openacc programming model using rcuda.
\newblock In {\em 2015 {IEEE} International Conference on Cluster Computing,
  {CLUSTER} 2015, Chicago, IL, USA, September 8-11, 2015\/} (2015), {IEEE}
  Computer Society, pp.~92--95.

\bibitem{farm}
{\sc Dragojevic, A., Narayanan, D., Castro, M., and Hodson, O.}
\newblock {FaRM}: Fast remote memory.
\newblock In {\em Proceedings of the 11th {USENIX} Symposium on Networked
  Systems Design and Implementation, {NSDI} 2014, Seattle, WA, USA, April 2-4,
  2014\/} (2014), R.~Mahajan and I.~Stoica, Eds., {USENIX} Association,
  pp.~401--414.

\bibitem{DBLP:conf/europar/DuatoIMPQS09}
{\sc Duato, J., Igual, F.~D., Mayo, R., Pe{\~{n}}a, A.~J.,
  Quintana{-}Ort{\'{\i}}, E.~S., and Silla, F.}
\newblock An efficient implementation of {GPU} virtualization in high
  performance clusters.
\newblock In {\em Euro-Par 2009 - Parallel Processing Workshops, HPPC,
  HeteroPar, PROPER, ROIA, UNICORE, VHPC, Delft, The Netherlands, August 25-28,
  2009, Revised Selected Papers\/} (2009), H.~Lin, M.~Alexander, M.~Forsell,
  A.~Kn{\"{u}}pfer, R.~Prodan, L.~Sousa, and A.~Streit, Eds., vol.~6043 of {\em
  Lecture Notes in Computer Science}, Springer, pp.~385--394.

\bibitem{duato2010modeling}
{\sc Duato, J., Pena, A.~J., Silla, F., Mayo, R., and Quintana-Ort{\i}, E.~S.}
\newblock Modeling the cuda remoting virtualization behaviour in high
  performance networks.
\newblock In {\em First Workshop on Language, Compiler, and Architecture
  Support for GPGPU\/} (2010).

\bibitem{DBLP:conf/ieeehpcs/DuatoPSMQ10}
{\sc Duato, J., Pe{\~{n}}a, A.~J., Silla, F., Mayo, R., and
  Quintana{-}Ort{\'{\i}}, E.~S.}
\newblock rcuda: Reducing the number of gpu-based accelerators in high
  performance clusters.
\newblock In {\em Proceedings of the 2010 International Conference on High
  Performance Computing {\&} Simulation, {HPCS} 2010, June 28 - July 2, 2010,
  Caen, France\/} (2010), W.~W. Smari and J.~P. McIntire, Eds., {IEEE},
  pp.~224--231.

\bibitem{DBLP:conf/icpp/DuatoPSMQ11}
{\sc Duato, J., Pe{\~{n}}a, A.~J., Silla, F., Mayo, R., and
  Quintana{-}Ort{\'{\i}}, E.~S.}
\newblock Performance of {CUDA} virtualized remote gpus in high performance
  clusters.
\newblock In {\em International Conference on Parallel Processing, {ICPP} 2011,
  Taipei, Taiwan, September 13-16, 2011\/} (2011), G.~R. Gao and Y.~Tseng,
  Eds., {IEEE} Computer Society, pp.~365--374.

\bibitem{DBLP:journals/concurrency/EilingBLM22}
{\sc Eiling, N., Baude, J., Lankes, S., and Monti, A.}
\newblock Cricket: {A} virtualization layer for distributed execution of {CUDA}
  applications with checkpoint/restart support.
\newblock {\em Concurr. Comput. Pract. Exp. 34}, 14 (2022).

\bibitem{huggingface}
{\sc Face, H.}
\newblock The ai community building the future.
\newblock \burl{https://huggingface.co}, 2023.

\bibitem{DBLP:conf/ipps/FinglerZYJWR22}
{\sc Fingler, H., Zhu, Z., Yoon, E., Jia, Z., Witchel, E., and Rossbach, C.~J.}
\newblock {DGSF:} disaggregated gpus for serverless functions.
\newblock In {\em 2022 {IEEE} International Parallel and Distributed Processing
  Symposium, {IPDPS} 2022, Lyon, France, May 30 - June 3, 2022\/} (2022),
  {IEEE}, pp.~739--750.

\bibitem{DBLP:conf/osdi/GaoNKCH0RS16}
{\sc Gao, P.~X., Narayan, A., Karandikar, S., Carreira, J., Han, S., Agarwal,
  R., Ratnasamy, S., and Shenker, S.}
\newblock Network requirements for resource disaggregation.
\newblock In {\em 12th {USENIX} Symposium on Operating Systems Design and
  Implementation, {OSDI} 2016, Savannah, GA, USA, November 2-4, 2016\/} (2016),
  K.~Keeton and T.~Roscoe, Eds., {USENIX} Association, pp.~249--264.

\bibitem{DBLP:conf/europar/GiuntaMAC10}
{\sc Giunta, G., Montella, R., Agrillo, G., and Coviello, G.}
\newblock A {GPGPU} transparent virtualization component for high performance
  computing clouds.
\newblock In {\em Euro-Par 2010 - Parallel Processing, 16th International
  Euro-Par Conference, Ischia, Italy, August 31 - September 3, 2010,
  Proceedings, Part {I}\/} (2010), P.~D'Ambra, M.~R. Guarracino, and D.~Talia,
  Eds., vol.~6271 of {\em Lecture Notes in Computer Science}, Springer,
  pp.~379--391.

\bibitem{guleria2019emf}
{\sc Guleria, A., Lakshmi, J., and Padala, C.}
\newblock Emf: Disaggregated gpus in datacenters for efficiency, modularity and
  flexibility.
\newblock In {\em 2019 IEEE International Conference on Cloud Computing in
  Emerging Markets (CCEM)\/} (2019), IEEE, pp.~1--8.

\bibitem{DBLP:conf/sigcomm/GuoWDSYPL16}
{\sc Guo, C., Wu, H., Deng, Z., Soni, G., Ye, J., Padhye, J., and Lipshteyn,
  M.}
\newblock {RDMA} over commodity ethernet at scale.
\newblock In {\em Proceedings of the {ACM} {SIGCOMM} 2016 Conference,
  Florianopolis, Brazil, August 22-26, 2016\/} (2016), M.~P. Barcellos,
  J.~Crowcroft, A.~Vahdat, and S.~Katti, Eds., {ACM}, pp.~202--215.

\bibitem{DBLP:conf/eurosys/GuptaGSKTTR09}
{\sc Gupta, V., Gavrilovska, A., Schwan, K., Kharche, H., Tolia, N., Talwar,
  V., and Ranganathan, P.}
\newblock Gvim: Gpu-accelerated virtual machines.
\newblock In {\em Proceedings of the 3rd {ACM} Workshop on System-level
  Virtualization for High Performance Computing, HPCVirt '09, Nuremburg,
  Germany, March 31, 2009\/} (2009), S.~L. Scott and G.~Vall{\'{e}}e, Eds.,
  {ACM}, pp.~17--24.

\bibitem{DBLP:conf/usenix/GuptaSTTR11}
{\sc Gupta, V., Schwan, K., Tolia, N., Talwar, V., and Ranganathan, P.}
\newblock Pegasus: Coordinated scheduling for virtualized accelerator-based
  systems.
\newblock In {\em 2011 {USENIX} Annual Technical Conference, Portland, OR, USA,
  June 15-17, 2011\/} (2011), J.~Nieh and C.~A. Waldspurger, Eds., {USENIX}
  Association.

\bibitem{DBLP:journals/corr/abs-2310-04648}
{\sc He, B., Zheng, X., Chen, Y., Li, W., Zhou, Y., Long, X., Zhang, P., Lu,
  X., Jiang, L., Liu, Q., Cai, D., and Zhang, X.}
\newblock Dxpu: Large scale disaggregated {GPU} pools in the datacenter.
\newblock {\em CoRR abs/2310.04648\/} (2023).

\bibitem{DBLP:journals/csur/HongSN17}
{\sc Hong, C., Spence, I. T.~A., and Nikolopoulos, D.~S.}
\newblock {GPU} virtualization and scheduling methods: {A} comprehensive
  survey.
\newblock {\em {ACM} Comput. Surv. 50}, 3 (2017), 35:1--35:37.

\bibitem{DBLP:conf/hotos/HuWWSBZKZCXZFS23}
{\sc Hu, C., Wang, C., Wang, S., Sun, N., Bao, Y., Zhao, J., Kashyap, S., Zuo,
  P., Chen, X., Xu, L., Zhang, Q., Feng, H., and Shan, Y.}
\newblock Skadi: Building a distributed runtime for data systems in
  disaggregated data centers.
\newblock In {\em Proceedings of the 19th Workshop on Hot Topics in Operating
  Systems, {HOTOS} 2023, Providence, RI, USA, June 22-24, 2023\/} (2023),
  M.~Schwarzkopf, A.~Baumann, and N.~Crooks, Eds., {ACM}, pp.~94--102.

\bibitem{DBLP:conf/usenix/KaliaKA16}
{\sc Kalia, A., Kaminsky, M., and Andersen, D.~G.}
\newblock Design guidelines for high performance {RDMA} systems.
\newblock In {\em 2016 {USENIX} Annual Technical Conference, {USENIX} {ATC}
  2016, Denver, CO, USA, June 22-24, 2016\/} (2016), A.~Gulati and
  H.~Weatherspoon, Eds., {USENIX} Association, pp.~437--450.

\bibitem{alpaserve}
{\sc Li, Z., Zheng, L., Zhong, Y., Liu, V., Sheng, Y., Jin, X., Huang, Y.,
  Chen, Z., Zhang, H., Gonzalez, J.~E., and Stoica, I.}
\newblock {AlpaServe}: Statistical multiplexing with model parallelism for deep
  learning serving.
\newblock In {\em 17th USENIX Symposium on Operating Systems Design and
  Implementation (OSDI 23)\/} (Boston, MA, July 2023), USENIX Association,
  pp.~663--679.

\bibitem{lim2014mica}
{\sc Lim, H., Han, D., Andersen, D.~G., and Kaminsky, M.}
\newblock Mica: A holistic approach to fast in-memory key-value storage.
\newblock In {\em Proceedings of the 11th USENIX Conference on Networked
  Systems Design and Implementation\/} (Berkeley, CA, USA, 2014), NSDI'14,
  USENIX Association, pp.~429--444.

\bibitem{connect7}
{\sc Mellanox}.
\newblock {ConnectX-7} product brief.
\newblock
  \burl{https://www.nvidia.com/content/dam/en-zz/Solutions/networking/ethernet-adapters/connectx-7-datasheet-Final.pdf},
  2022.

\bibitem{ray}
{\sc Moritz, P., Nishihara, R., Wang, S., Tumanov, A., Liaw, R., Liang, E.,
  Elibol, M., Yang, Z., Paul, W., Jordan, M.~I., and Stoica, I.}
\newblock Ray: A distributed framework for emerging {AI} applications.
\newblock In {\em 13th USENIX Symposium on Operating Systems Design and
  Implementation (OSDI 18)\/} (Carlsbad, CA, Oct. 2018), USENIX Association,
  pp.~561--577.

\bibitem{narayanan2021efficient}
{\sc Narayanan, D., Shoeybi, M., Casper, J., LeGresley, P., Patwary, M.,
  Korthikanti, V., Vainbrand, D., Kashinkunti, P., Bernauer, J., Catanzaro, B.,
  et~al.}
\newblock Efficient large-scale language model training on gpu clusters using
  megatron-lm.
\newblock In {\em Proceedings of the International Conference for High
  Performance Computing, Networking, Storage and Analysis\/} (2021), pp.~1--15.

\bibitem{DBLP:conf/sosp/NgD023}
{\sc Ng, K. K.~W., Demoulin, H.~M., and Liu, V.}
\newblock Paella: Low-latency model serving with software-defined {GPU}
  scheduling.
\newblock In {\em Proceedings of the 29th Symposium on Operating Systems
  Principles, {SOSP} 2023, Koblenz, Germany, October 23-26, 2023\/} (2023),
  J.~Flinn, M.~I. Seltzer, P.~Druschel, A.~Kaufmann, and J.~Mace, Eds., {ACM},
  pp.~595--610.

\bibitem{dgx100}
{\sc NVIDIA}.
\newblock {NVIDIA DGX Platform}.
\newblock \burl{https://www.nvidia.com/en-us/data-center/dgx-platform/}, 2024.

\bibitem{rdma-perf}
{\sc OFED}.
\newblock {Open Fabrics Enterprise Distribution (OFED) Performance Tests}.
\newblock \burl{https://github.com/linux-rdma/perftest}, 2024.

\bibitem{pope2023efficiently}
{\sc Pope, R., Douglas, S., Chowdhery, A., Devlin, J., Bradbury, J., Heek, J.,
  Xiao, K., Agrawal, S., and Dean, J.}
\newblock Efficiently scaling transformer inference.
\newblock {\em Proceedings of Machine Learning and Systems 5\/} (2023).

\bibitem{pytorch-dataloader}
{\sc Pytorch}.
\newblock Datasets and dataloaders.
\newblock
  \burl{https://pytorch.org/tutorials/beginner/basics/data_tutorial.html},
  2024.

\bibitem{torch-elastic}
{\sc Pytorch}.
\newblock Torchelastic.
\newblock \burl{https://pytorch.org/elastic/latest/}, 2024.

\bibitem{DBLP:conf/osdi/RuanSAB20}
{\sc Ruan, Z., Schwarzkopf, M., Aguilera, M.~K., and Belay, A.}
\newblock {AIFM:} high-performance, application-integrated far memory.
\newblock In {\em 14th {USENIX} Symposium on Operating Systems Design and
  Implementation, {OSDI} 2020, Virtual Event, November 4-6, 2020\/} (2020),
  {USENIX} Association, pp.~315--332.

\bibitem{DBLP:conf/osdi/ShanHCZ18}
{\sc Shan, Y., Huang, Y., Chen, Y., and Zhang, Y.}
\newblock Legoos: {A} disseminated, distributed {OS} for hardware resource
  disaggregation.
\newblock In {\em 13th {USENIX} Symposium on Operating Systems Design and
  Implementation, {OSDI} 2018, Carlsbad, CA, USA, October 8-10, 2018\/} (2018),
  A.~C. Arpaci{-}Dusseau and G.~Voelker, Eds., {USENIX} Association,
  pp.~69--87.

\bibitem{DBLP:journals/tc/ShiCSL12}
{\sc Shi, L., Chen, H., Sun, J., and Li, K.}
\newblock vcuda: Gpu-accelerated high-performance computing in virtual
  machines.
\newblock {\em {IEEE} Trans. Computers 61}, 6 (2012), 804--816.

\bibitem{DBLP:journals/corr/abs-2202-07848}
{\sc Shukla, D., Sivathanu, M., Viswanatha, S., Gulavani, B.~S., Nehme, R.,
  Agrawal, A., Chen, C., Kwatra, N., Ramjee, R., Sharma, P., Katiyar, A., Modi,
  V., Sharma, V., Singh, A., Singhal, S., Welankar, K., Xun, L., Anupindi, R.,
  Elangovan, K., Rahman, H., Lin, Z., Seetharaman, R., Xu, C., Ailijiang, E.,
  Krishnappa, S., and Russinovich, M.}
\newblock Singularity: Planet-scale, preemptive and elastic scheduling of {AI}
  workloads.
\newblock {\em CoRR abs/2202.07848\/} (2022).

\bibitem{cachecloud}
{\sc Thomas, S., Voelker, G.~M., and Porter, G.}
\newblock Cachecloud: Towards speed-of-light datacenter communication.
\newblock In {\em 10th {USENIX} Workshop on Hot Topics in Cloud Computing,
  HotCloud 2018, Boston, MA, USA, July 9, 2018\/} (2018), G.~Ananthanarayanan
  and I.~Gupta, Eds., {USENIX} Association.

\bibitem{DBLP:conf/usenix/TsaiSZ20}
{\sc Tsai, S., Shan, Y., and Zhang, Y.}
\newblock Disaggregating persistent memory and controlling them remotely: An
  exploration of passive disaggregated key-value stores.
\newblock In {\em 2020 {USENIX} Annual Technical Conference, {USENIX} {ATC}
  2020, July 15-17, 2020\/} (2020), A.~Gavrilovska and E.~Zadok, Eds., {USENIX}
  Association, pp.~33--48.

\bibitem{tsai2017lite}
{\sc Tsai, S.-Y., and Zhang, Y.}
\newblock Lite kernel rdma support for datacenter applications.
\newblock In {\em Proceedings of the 26th Symposium on Operating Systems
  Principles\/} (New York, NY, USA, 2017), SOSP '17, ACM, pp.~306--324.

\bibitem{vmware-bit}
{\sc VMWare}.
\newblock Mware vsphere bitfusion.
\newblock
  \burl{https://docs.vmware.com/cn/VMware-vSphere-Bitfusion/index.html}, 2023.

\bibitem{DBLP:conf/osdi/WangMLLRNBNKX20}
{\sc Wang, C., Ma, H., Liu, S., Li, Y., Ruan, Z., Nguyen, K., Bond, M.~D.,
  Netravali, R., Kim, M., and Xu, G.~H.}
\newblock Semeru: {A} memory-disaggregated managed runtime.
\newblock In {\em 14th {USENIX} Symposium on Operating Systems Design and
  Implementation, {OSDI} 2020, Virtual Event, November 4-6, 2020\/} (2020),
  {USENIX} Association, pp.~261--280.

\bibitem{drtm-h}
{\sc Wei, X., Dong, Z., Chen, R., and Chen, H.}
\newblock Deconstructing {RDMA-enabled} distributed transactions: Hybrid is
  better!
\newblock In {\em 13th USENIX Symposium on Operating Systems Design and
  Implementation (OSDI 18)\/} (Carlsbad, CA, Oct. 2018), USENIX Association,
  pp.~233--251.

\bibitem{DBLP:journals/corr/abs-2306-03622}
{\sc Yu, M., Wang, A., Chen, D., Yu, H., Luo, X., Li, Z., Wang, W., Chen, R.,
  Nie, D., and Yang, H.}
\newblock Faaswap: Slo-aware, gpu-efficient serverless inference via model
  swapping.
\newblock {\em CoRR abs/2306.03622\/} (2023).

\bibitem{DBLP:journals/tpds/ZhangYQYG14}
{\sc Zhang, C., Yao, J., Qi, Z., Yu, M., and Guan, H.}
\newblock vgasa: Adaptive scheduling algorithm of virtualized {GPU} resource in
  cloud gaming.
\newblock {\em {IEEE} Trans. Parallel Distributed Syst. 25}, 11 (2014),
  3036--3045.

\bibitem{DBLP:conf/sosp/ZhangMHLCDDJMW23}
{\sc Zhang, M., Ma, T., Hua, J., Liu, Z., Chen, K., Ding, N., Du, F., Jiang,
  J., Ma, T., and Wu, Y.}
\newblock Partial failure resilient memory management system for (cxl-based)
  distributed shared memory.
\newblock In {\em Proceedings of the 29th Symposium on Operating Systems
  Principles, {SOSP} 2023, Koblenz, Germany, October 23-26, 2023\/} (2023),
  J.~Flinn, M.~I. Seltzer, P.~Druschel, A.~Kaufmann, and J.~Mace, Eds., {ACM},
  pp.~658--674.

\bibitem{racehashing}
{\sc Zuo, P., Sun, J., Yang, L., Zhang, S., and Hua, Y.}
\newblock One-sided rdma-conscious extendible hashing for disaggregated memory.
\newblock In {\em 2021 {USENIX} Annual Technical Conference ({USENIX} {ATC}
  21)\/} (July 2021), {USENIX} Association, pp.~15--29.

\end{thebibliography}
}

\clearpage

\end{document}